\documentclass{article}

\usepackage[utf8]{inputenc}
\usepackage[english]{babel}
\usepackage{authblk}
\usepackage{amssymb,amsmath,amsthm,twoopt,xargs,mathtools}
\usepackage{times,dsfont,ifthen}
\usepackage{fancyhdr,xcolor,breakcites,booktabs}

\title{Backward importance sampling for online estimation of state space models}
\date{}

\author[$\dag\,\amalg$]{Alice Martin\footnote{This action benefited from the support of the Chair « New Gen RetAIl » led by l’X – École polytechnique and the Fondation de l’École polytechnique, sponsored by CARREFOUR}}
\author[$\wr$]{Marie-Pierre \'Etienne}
\author[$\star$]{Pierre Gloaguen}
\author[$\dag$]{Sylvain Le Corff}
\author[$\ddag$]{Jimmy Olsson}

\affil[$\dag$]{{\small Samovar, T\'el\'ecom SudParis, d\'epartement CITI, TIPIC, Institut Polytechnique de Paris, Palaiseau.}}
\affil[$\amalg$]{{\small CMAP, \'Ecole Polytechnique, Institut Polytechnique de Paris, Palaiseau.}}
\affil[$\wr$]{{\small Agrocampus Ouest, CNRS, IRMAR - UMR 6625, F-35000 Rennes.}}
\affil[$\star$]{{\small AgroParisTech, UMR MIA 518.}}
\affil[$\ddag$]{{\small Department of Mathematics, KTH Royal Institute of Technology, Stockholm.}}

\usepackage{geometry}
\newcommand{\gpeUB}{\overline{m}_{\parvec}}
\newcommand{\gpeLB}{\underline{m}_{\parvec}}

\def\Xset{\mathbb{R}^d}

\newcommand{\md}[1]{g_{#1}}

\newcommand{\logllh}[1]{\ell_{#1}}
\newcommand{\llh}[1]{\mathsf{L}_{#1}}

\newcommandx\filtderiv[2][1=]{
\ifthenelse{\equal{#1}{}}
	{\eta_{#2}}
	{\eta_{#2}^\N}
}
\newcommand{\pred}[1]{\pi_{#1}}
\newcommand{\parvec}{\theta}
\newcommand{\parspace}{\Theta}
\newcommand{\tstatletter}{\kernel{T}}

\newcommandx\tstat[2][1=]{
\ifthenelse{\equal{#1}{}}
	{\tstatletter_{#2}}
	{\tau_{#2}^{#1}}
}
\newcommandx\tstathat[2][1=]{
\ifthenelse{\equal{#1}{}}
	{\tstatletter_{#2}}
	{\widehat{\tau}_{#2}^{#1}}
}
\newcommand{\af}[1]{h_{#1}} 
\newcommand{\deriv}{\nabla_{\parvec}}

\newcommand{\kernel}[1]{\mathbf{#1}}

\newcommand{\set}[1]{\mathsf{#1}}

\newcommandx{\bk}[2][1=]{ 
\ifthenelse{\equal{#1}{}}
{\overleftarrow{\kernel{Q}}_{#2}}
{\overleftarrow{\kernel{Q}}_{#2}^{#1}}
}

\newcommandx{\bkhat}[2][1=]{ 
\ifthenelse{\equal{#1}{}}
{\widehat{\kernel{Q}}_{#2}}
{\widehat{\kernel{Q}}_{#2}^{#1}}
}

\newcommand{\hd}[1]{q_{#1}} 
\newcommand{\hdhat}[1]{\widehat{q}_{#1}}

\newcommand{\addf}[1]{\termletter_{#1}}

\newcommand{\termletter}{\tilde{h}}
\newcommand{\N}{N}

\newcommandx{\K}[1][1=]{
\ifthenelse{\equal{#1}{}}{{\kletter}}{{\widetilde{\N}^{#1}}}}
\newcommand{\hkup}{\bar{\varepsilon}}
\newcommand{\bi}[3]{J_{#1}^{(#2, #3)}}

\newcommand{\kletter}{\widetilde{\N}}

\def\1{\mathds{1}}
\def\pE{\mathbb{E}}

\newcommand{\esssup}[2][]
{\ifthenelse{\equal{#1}{}}{\left\| #2 \right\|_\infty}{\left\| #2 \right\|^2_{\infty}}}

\newcommand{\rset}{\ensuremath{\mathbb{R}}}

\newcommand{\kiss}[3][]
{\ifthenelse{\equal{#1}{}}{r_{#2|#3}}
{\ifthenelse{\equal{#1}{fully}}{r^{\star}_{#2|#3}}
{\ifthenelse{\equal{#1}{smooth}}{\tilde{r}_{#2|#3}}{\mathrm{erreur}}}}}

\newcommand{\chunk}[4][]%
{\ifthenelse{\equal{#1}{}}{\ensuremath{{#2}_{#3:#4}}}{\ensuremath{#2^#1}_{#3:#4}}
}

\newcommand{\kissforward}[3][]
{\ifthenelse{\equal{#1}{}}{p_{#2}}
{\ifthenelse{\equal{#1}{fully}}{p^{\star}_{#2}}
{\ifthenelse{\equal{#1}{smooth}}{\tilde{r}_{#2}}{\mathrm{erreur}}}}}

\newcommand{\instrpostaux}[1]{\ensuremath{\upsilon_{#1}}}
\newcommandx\post[2][1=]{
\ifthenelse{\equal{#1}{}}
	{\phi_{#2}}
	{\phi_{#2}^\N}
}

\newcommandx\posthat[2][1=]{
\ifthenelse{\equal{#1}{}}
	{\widehat{\phi}_{#2}}
	{\widehat{\phi}_{#2}^\N}
}

\newcommand{\adjfunc}[4][]
{\ifthenelse{\equal{#1}{}}{\ifthenelse{\equal{#4}{}}{\vartheta_{#2|#3}}{\vartheta_{#2|#3}(#4)}}
{\ifthenelse{\equal{#1}{smooth}}{\ifthenelse{\equal{#4}{}}{\tilde{\vartheta}_{#2|#3}}{\tilde{\vartheta}_{#2|#3}(#4)}}
{\ifthenelse{\equal{#1}{fully}}{\ifthenelse{\equal{#4}{}}{\vartheta^\star_{#2|#3}}{\vartheta^\star_{#2|#3}(#4)}}{\mathrm{erreur}}}}}

\newcommand{\XinitIS}[2][]
{\ifthenelse{\equal{#1}{}}{\ensuremath{\rho_{#2}}}{\ensuremath{\check{\rho}_{#2}}}}

\newcommand{\rmd}{\ensuremath{\mathrm{d}}}
\newcommand{\eqdef}{\ensuremath{:=}}
\newcommand{\eqsp}{\;}
\newcommand{\ewght}[2]{\ensuremath{\omega_{#1}^{#2}}}

\newcommand{\epart}[2]{\ensuremath{\xi_{#1}^{#2}}}
\newcommand{\filt}[2][]%
{%
\ifthenelse{\equal{#1}{}}{\ensuremath{\phi_{#2}}}{\ensuremath{\phi_{#1,#2}}}%
}
\newcommand{\Xinit}{\ensuremath{\chi}}
\newcommand{\sumwght}[2][]{%
\ifthenelse{\equal{#1}{}}{\ensuremath{\Omega_{#2}}}{\ensuremath{\Omega_{#2}^{(#1)}}}}
\newcommand{\sumwghthat}[2][]{%
\ifthenelse{\equal{#1}{}}{\ensuremath{\widehat{\Omega}_{#2}}}{\ensuremath{\widehat{\Omega}_{#2}^{(#1)}}}}

\newcounter{hypH}
\newenvironment{hypH}{\refstepcounter{hypH}\begin{itemize}
\item[{\bf H\arabic{hypH}}]}{\end{itemize}}

\newcommand{\marginalset}{\mathsf{U}}

\newcommand{\kernelmarg}{\mathbf{R}}

\newcommand{\qg}[1]{\ell_{#1}}
\newcommand{\hatqg}[1]{\mathsf{\ell}_{#1}}

\newcommand{\W}{\mathbf{W}}

\newcounter{example}[section]
\newenvironment{example}[1][]{\refstepcounter{example}\par\medskip
   \noindent \textbf{Example~\theexample:} \textit{#1} \text \rmfamily}{\medskip}

\begin{document}

\maketitle

\begin{abstract}
This paper proposes a new Sequential Monte Carlo algorithm to perform online estimation in the context of state space models when either the transition density of the latent state or the conditional likelihood of an observation given a state is intractable. 
In this setting, obtaining low variance estimators of expectations under the posterior distributions of the unobserved states given the observations is a challenging task. 
Following recent theoretical results for pseudo-marginal sequential Monte Carlo smoothers, a pseudo-marginal backward importance sampling step is introduced to estimate such expectations. 
This new step allows to reduce very significantly the computational time  of the existing numerical solutions based on an acceptance-rejection procedure for similar performance, and to broaden the class of eligible models for such methods. 
For instance, in the context of multivariate stochastic differential equations, the proposed algorithm makes use of unbiased estimates of the unknown transition densities under much weaker assumptions than standard alternatives.
 The performance of this estimator is assessed for high-dimensional discrete-time latent data models,  for recursive maximum likelihood estimation in the context of partially observed diffusion process, and  in the case of a bidimensional partially observed stochastic Lotka-Volterra model. 
\end{abstract}

\section{Introduction}
\label{sec:intro}
Latent data models are widely used in time series and sequential data analysis across a wide range of applied science
and engineering domains such as movement ecology \cite{michelot2016movehmm}, energy consumptions modelling \cite{candanedo2017methodology}, genomics \cite{yau2011bayesian, gassiat2016inference, wang2017variational},  target tracking \cite{sarkka2007rao}, enhancement and segmentation of speech and audio signals \cite{rabiner1989tutorial}, see also \cite{sarkka2013bayesian, douc2014nonlinear, zucchini2017hidden} and the numerous references therein.  Performing maximum likehood estimation (MLE) for instance with the Expectation Maximization (EM) algorithm  \cite{dempster1977maximum} or a stochastic gradient ascent (\cite{cappe2005inference} in the case of HMMs) is a challenging task.  Both approaches involve conditional distributions of sequences of hidden states given the observation record (the \textit{smoothing} distribution), which are not available explicitly.



Markov chain Monte Carlo (MCMC) and sequential Monte Carlo (SMC) methods (also known as particle filters or smoothers) are widespread solutions to propose consistent estimators of such distributions. 
Among SMC methods, algorithms have been designed in the last decades to solve the smoothing problem, such as the Forward Filtering Backward Simulation algorithm \cite{douc2011sequential}  or two-filter based approaches \cite{briers2010smoothing, fearnhead2010sequential, nguyen2017two}. 
These approaches, which come with strong theoretical guarantees (\cite{delmoral2010backward, douc2011sequential, dubarry2013nonasymptotic, gerber2017convergence}), require the time horizon and all observations to be available  to initialize a backward information filter, and, thus, perform the smoothing.
The particle-based rapid incremental smoother \cite{olsson2017efficient} is an online version of forward-backward procedures, specifically designed to approximate conditional expectations of additive functionals. 
This algorithm relies on a backward sampling step performed on the fly thanks to the well known acceptance rejection sampling. 
This online smoother was proven to be strongly consistent, asymptotically normal, and with a control of the asymptotic variance, when it is performed together with the vanilla bootstrap filter \cite{gordon1993novel}. 
In \cite{olsson2020particle}, the authors show how this algorithm can be used to performed recursive maximum likelihood in state space models. This approach relies on the necessity to upper bound the transition density of the hidden signal, as it is required to perform acceptance rejection sampling.

Moreover, a pivotal step of all SMC approaches is the evaluation of this transition density and of the density of the conditional distribution of an observation given the corresponding latent state (the marginal conditional likelihood).
In many practical settings, though, no closed-form expressions of these distributions are available: for instance, in the case of partially observed diffusions \cite{andersson2017unbiased,fearnhead2017continuous} or in the context of  approximate Bayesian computation smoothing \cite{martin:jasra:singh:whiteley:delmoral:maccoy:2014}.
 A first step to bypass this shortcoming was proposed in  \cite{fearnhead2010random}. 
The authors proposed an important contribution by showing that it is possible to implement importance sampling and filtering recursions, when the unavailable importance weights are replaced by random estimators.
Standard data augmentation schemes were then used to extend this random-weight particle filter to provide new inference procedures for instance for partially observed diffusion models \cite{yonekura:beskos:2020}. 

More recently, the online algorithm of \cite{olsson2017efficient} was extended to this setting for partially observed diffusion processes by \cite{gloaguen2018online}.  Then, \cite{gloaguen2021pseudo} introduced a pseudo-marginal online smoother to approximate conditional expectations of additive functionals of the hidden states in a very general setting: the user can only evaluate (possibly biased) approximations of the transition density and of the marginal conditional likelihood. 
The online algorithm of \cite{gloaguen2021pseudo} may be used to approximate expectations of additive functionals under the smoothing distributions by processing the data stream online.
However, as with the PaRIS algorihm, when using this pseudo-marginal approach where transition densities are intractable, the user needs to sample exactly from the associated pseudo-marginal backward kernel. 
This step is again done by rejection sampling, and therefore requires that the estimate of the transition density and of the marginal conditional  likelihood  are almost surely positive and upper bounded.  
In practice, these assumptions are very restrictive. 
For instance, in the context of diffusion processes, they narrow the possible models to the class of diffusions satisfying the Exact algorithm conditions of \cite{beskos2006retrospective}, for which  General Poisson Estimators (GPEs) \cite{fearnhead2008particle} already lead to eligible unbiased estimators.  

In this paper, a new procedure is introduced to replace the backward acceptance-rejection step of the PaRIS and the pseudo marginal PaRIS algorithms by a backward importance sampling estimate.  
It leads to a smoothing algorithm that only requires an almost surely positive estimator of the unknown transition or observation density, and therefore extends widely the class of models for which these online smoothers can be designed. 
In the general case where only signed estimates can be obtained, we propose to use  Wald's trick, ensuring positiveness. 
 In the context of partially observed diffusion processes, for instance,  we show that combining Wald's trick to the  parametrix estimators of \cite{andersson2017unbiased} and \cite{fearnhead2017continuous} leads to a highly generic algorithm that can be applied to a wide class of models, for which no low variance smoother existed so far.

The paper is organized as follows. Section~\ref{sec:model} displays the latent data models and the main objectives considered in this paper. Then, Section~\ref{sec:method} details online pseudo marginal sequential Monte Carlo algorithms and Section~\ref{sec:backwardis} our proposed algorithm. Section~\ref{sec:application}  provides  extensive numerical experiments to illustrate the performance of our approach. The empirical results of this section can be summarised as follows.
\begin{itemize}
\item The proposed approach can be used for any latent data models such as hidden Markov models, or recurrent neural networks with unobserved latent states. Even when the transition densities are available, we show empirically that our backward importance sampling is a computationally efficient solution to solve the online smoothing problem (Section \ref{sec:simu:RNN}).
\item We show that the proposed approach outperforms the existing acceptance rejection method in terms of computational efficiency 
(Section \ref{sec:simu:SINE}). 
\item We show how the proposed method allows for efficient online recursive maximum likelihood in the context of partially observed diffusion processes (Section \ref{sec:simu:tangent:filter}). 
\item When considering the pseudo-marginal approach, we extend the use of Wald's trick to the backward kernel, and therefore show that our approach can be used in cases where the estimators of the unknown densities  are not positive by construction.
\item  We perform sequential Monte Carlo smoothing in models for which no solutions were proposed in the literature to the best of our knowledge, such as multivariate partially observed diffusion processes (Section \ref{sec:simu:LV}).
\end{itemize}

\section{Model and objectives}
\label{sec:model}

Let $\parvec$ be a parameter lying in a $\Theta\subset \rset^q$ and consider a  \textit{state space model} where the hidden Markov chain  in $\rset^d$ is denoted by $(X_k)_{k\geqslant 0}$. The distribution of $X_0$ has density $\chi$ with respect to the Lebesgue measure and for all $0\leqslant k \leqslant n-1$, the conditional distribution of $X_{k+1} $ given $X_{0:k}$ has density $\hd{k+1;\parvec}(X_{k},\cdot)$, where $a_{u:v}$ is a short-hand notation for $(a_u,\ldots,a_v)$. 
It is assumed that this state  is partially observed  through an observation process $(Y_k)_{0\leqslant k \leqslant n}$ taking values in $\rset^m$. 
For all $0\leqslant k \leqslant n$, the distribution of $Y_k$ given $X_{0:n}$ depends on $X_k$ only and has density $\md{k;\parvec}(X_k,\cdot)$ with respect to the Lebesgue measure. 
In this context, for any pair of indexes $0\leqslant\leqslant k_1 \leqslant k_2 \leqslant n$, we define the \textit{joint smoothing distribution} as the conditional law of $X_{k_1:k_2}$ given $Y_{0:n}$. 
In this framework, the likelihood of the observations $\llh{n,\parvec}(Y_{0:n})$, which is  in general intractable, is
$$
\llh{n,\parvec}(Y_{0:n})  = \int \chi(x_0)\md{0;\parvec}(x_{0},Y_{0})\prod_{k=0}^{n-1}\qg{k;\parvec}(x_{k},x_{k+1})\rmd x_{0:n}\eqsp,
$$
 where, for all $0\leqslant k \leqslant n$ and all $\parvec\in\parspace$,
\begin{equation}
\label{eq:def:elln}
\qg{k;\parvec}(x_{k},x_{k+1}) = \hd{k+1;\parvec}(x_{k}, x_{k+1}, Y_{k + 1})\md{k+1;\parvec}(x_{k+1},Y_{k+1})\eqsp.
\end{equation}
In a large variety of situations, \eqref{eq:def:elln} cannot be evaluated pointwise  (see models of sections \ref{sec:simu:SINE} and \ref{sec:simu:LV}), and we assume in this paper that we have an estimate of this quantity (see assumption \textbf{H\ref{assum:unbiased}} in Section~\ref{sec:method}).
Note that to avoid future cumbersome expressions, the dependency of the key quantity $\qg{k;\parvec}(\cdot)$ on the observations is implicit, as we always work conditionnaly to the observations. In this paper, we propose an algorithm to compute \textit{smoothing expectations of additive functionals}. 
Namely, we aim at computing: 
$$
\pE \left[\af{0:n}(X_{0:n})\middle | Y_{0:n}\right]\eqsp,
$$
where $\af{0:n}$ is an \textit{additive functional}, \textit{i.e.} a function from $\rset^{d \times (n + 1)}$ to $\rset^{d'}$ satisfying:
\begin{equation}
\label{eq:additive:functional}
\af{0:n}: x_{0:n} \mapsto \sum_{k=0}^{n-1}\addf{k}(x_{k},x_{k+1})\eqsp,
\end{equation}
where $\addf{k}:\rset^{d} \times \rset^{d}\to\rset^{d'}$.
Such expectations are the keystones of many common inference problems in state space models.

\begin{example}[State estimation.]
\label{ex:state:tracking}
Suppose that the model parameter $\theta$ is known, a common objective is to recover the underlying signal $X_{k^*}$ for some index $0\leqslant k^* \leqslant n$ given the observations $Y_{0:n}$. A standard estimator is $\pE[X_{k^*}\vert Y_{0:n}]$, which is a particular instance of our problem with $\addf{k}(x_k, x_{k+1}) = x_k$ if $k = k^*$ and 0 otherwise. 
\end{example}

\begin{example}[EM algorithm.]
\label{ex:em:algorithm}
In the usual case when $\theta$ is unknown, the maximum likelihood estimator is $\widehat \parvec = \mathrm{argmax}_{\parvec\in\parspace}\eqsp\llh{n,\parvec}(Y_{0:n})$. Expectation Maximization based algorithms \cite{dempster1977maximum}  are appealing solutions to obtain an estimator of $\hat \parvec$.
The pivotal concept of the EM algorithm is that the intermediate quantity defined by
\begin{equation*}
\parvec\mapsto Q(\parvec,\parvec') = \pE_{\parvec'}\left[\sum_{k=0}^{n-1} \log \qg{k;\parvec}(X_{k}, X_{k+1})\middle | Y_{0:n}\right] 
\end{equation*}
may be used as a surrogate for $\llh{n}(\parvec)$ in the maximization procedure,  where $\pE_{\parvec'}$ is the expectation under the joint distribution of the latent states and the observations when the model is parameterized by $\parvec'$. 
Again, this inference setting is a special case of our framework where $\addf{k}(x_k, x_{k+1}) = \log \qg{k;\parvec}(x_{k}, x_{k+1})$.

\end{example}

\begin{example}[Fisher's identity and online gradient ascent.]
An alternative to the EM algorithm is to maximize the loglikelihood through gradient based methods.
Indeed, in state space models, under some regularity conditions (see \cite{cappe2005inference}, Chapter~10), the gradient of the log likelihood can be obtained thanks to Fisher's identity:
$$\deriv \log \llh{n}(\parvec) = \pE_\theta\left[\sum_{k=0}^{n-1} \deriv \log \qg{k;\parvec}(X_{k},X_{k+1})\middle | Y_{0:n}\right],$$
which relies on the expectation of a smoothing additive functional.
It has been noted (see \cite{cappe2005inference}, chapter 10, or \cite{olsson2020particle} how this identity, coupled with the smoothing recursions of Section \ref{sec:smoothing}, can lead to an online gradient ascent, that provides an online estimate of the MLE.
An extension of this method will be illustrated in Section~\ref{sec:simu:tangent:filter} in a challenging setting where the transition density cannot be evaluated.
\end{example}

\section{Online sequential Monte Carlo smoothing}
\label{sec:method}

\subsection{Backward statistic for online smoothing}

In this section, the parameter $\parvec$ is dropped from notation for a better clarity. For all pair of integers $0\leq k \leq k' \leq n$, and all measurable function $h$ on $\rset^{d \times (k' - k + 1)}$, the expectation with respect to the joint smoothing distribution is denoted by:
$$
\post{k:k' \mid n} \left[ h \right] := \pE\left[ h(X_{k:k'}) \middle | Y_{0:n} \right]\eqsp .
$$
The special case where $k = k' = n$ refers to \textit{filtering distribution} and we write $\post{k} = \post{k:k \mid k}$.
A pivotal quantity to estimate $\post{0:n \mid n}[\af{0:n}]$ is the \textit{backward statistic}:
\begin{equation}
\label{eq:T:stat}
\tstat{k}[\af{0:k}](X_k) = \pE \left[\af{0:k}(X_{0:k        }) \middle | X_k, Y_{0:k}\right],~1\leqslant k \leqslant n,~\tstat{0} = 0\eqsp.
\end{equation}
Note that for each $k$ this statistic is a function of $X_k$, and is defined relatively to the functional of interest $\af{0:n}$.
For additive functionals, this statistic satisfies the two following key identities, for all $1\leqslant k \leqslant n$:
\begin{align}
\post{0:n \mid n}[\af{0:n}] &= \post{n}\left[\tstat{n}[\af{0:n}] \right], \label{eq:property:filt:smooth}\\
\tstat{k}\left[\af{0:k}\right](X_{k}) &= \pE\left[\left.\tstat{k-1}\left[\af{0:(k-1)}\right]\left(X_{k-1}\right) + \addf{k-1}\left(X_{k-1},X_{k}\right)\right|X_{k},Y_{0:k-1} \right]\eqsp, \label{eq:property:backward}
\end{align} 
where $\addf{k-1}$ is the function defined in \eqref{eq:additive:functional}.
Property \eqref{eq:property:filt:smooth} essentialy tells us that the target is the filtering expectation of a well chosen statistic, while property \eqref{eq:property:filt:smooth} provides a recursion to compute these statistics. 
These two properties suggest an \textit{online} procedure to solve the online smoothing problem. Starting at time 0, at each step $k$, this procedure aims at (i) computing the filtering distribution and (ii) computing the backward statistics. 
Following \cite{fearnhead2008particle, olsson2011particle, gloaguen2018online, gloaguen2021pseudo}, we do not assume that  \eqref{eq:def:elln} can be evaluated pointwise.
We assume that there exists an estimator, relying on some random variable on a general state space $(\marginalset,\mathcal{B}(\marginalset))$ such that the following assumption holds.
\begin{hypH} 
\label{assum:unbiased}
For all $\parvec \in\parspace$ and $k\geqslant 0$, there exists a Markov kernel on $(\Xset\times\Xset,\mathcal{B}(\marginalset))$ with density $\kernelmarg_{k;\parvec}$ with respect to a reference measure $\mu$ on a general state space $(\marginalset,\mathcal{B}(\marginalset))$,  and a positive mapping $\hatqg{k;\parvec}\langle \cdot\rangle$ on $\Xset\times\Xset\times\marginalset$ such that, for all $(x,x')\in\Xset \times \Xset$,
\begin{equation*}
\int \kernelmarg_{k;\parvec}(x,x';z)\hatqg{k;\parvec}\langle z\rangle(x,x')\mu(\rmd z) =  \qg{k;\parvec}(x,x')\eqsp.
\end{equation*}
\end{hypH}
This setup, known as \textit{pseudo marginalisation} is based on the plug-in principle, as a pointwise estimate of $\qg{k;\parvec}(x,x')$ can be obtained by generating $\zeta$ from $\kernelmarg_{k;\parvec}(x,x';\rmd z)$ and computing the statistic $\hatqg{k;\parvec}\langle \zeta\rangle(x,x')$. 
Its use in Monte Carlo methods, and the related theoretical guarantees, have been studied in the context of MCMC \cite{andrieu2009pseudo}, and more recently, of SMC \cite{gloaguen2021pseudo}.

\subsubsection*{Recursive maximum likelihood}

An appealing application for online smoothing is the context of recursive maximum likelihood, i.e., where new observations are used only once to update the estimator of the unknown parameter $\parvec$.
Following \cite{legland1997recursive}, the idea is the build a sequence $\left\lbrace\parvec_k\right\rbrace_{k\geq 0}$ as follows. First, set the initial value of the parameter estimate: $\parvec_0$. Then, for each new observation $Y_{k},~k\geqslant 1$, define
$$
\theta_{k} = \theta_{k-1} + \gamma_k \deriv \logllh{\parvec}(Y_k \mid Y_{1:k - 1}) \eqsp,
$$
where $\logllh{\parvec}(Y_k \mid Y_{1:k - 1})$ is the log likelihood for the new observation given all the past, and $\left\lbrace\gamma_k\right\rbrace_{k\geqslant}$ are positive step sizes such that $\sum_{k \geqslant 1}\gamma_k = \infty$ and $\sum_{k \geqslant 1}\gamma_k^2 < \infty$. The practical implementation of such an update relies on the following identity:
\begin{equation}
\deriv \logllh{\parvec}(Y_k \mid Y_{1:k - 1})  
=  \frac{\pred{k;\parvec}[\deriv \md{k;\parvec}] + \filtderiv{k;\parvec}[\md{k;\parvec}]}{\pred{k;\parvec}[\md{k;\parvec}]}\eqsp,
\label{eq:online:gradient}
\end{equation}
where $\pred{k;\parvec} = \phi_{k;\parvec \mid k - 1}$ is the predictive distribution  and 
$$\filtderiv{k;\parvec}[\md{k;\parvec}] = \post{0:k;\parvec \mid k - 1} [\af{0:k;\parvec} \md{k;\parvec}] - \pred{k;\parvec}[\md{k;\parvec}] \times \post{0:k;\parvec \mid k - 1 }[\af{0:k}]\eqsp,$$
with
\begin{equation}
\label{eq:complete:gradient:log}
\af{0:k}(x_{0:k}) =  \sum_{j = 0}^{k - 1} \deriv\log \qg{j, \parvec}(x_j,x_{j+1})\eqsp.
\end{equation}
The signed measure $\filtderiv{k;\parvec}$ is known as the \textit{tangent filter}, see \cite[Chapter~10]{cappe2005inference}, \cite{delmoral2015uniform} or \cite{olsson2020particle}. 
Using the tower property  and the backward decomposition \eqref{eq:property:backward} yields
\begin{equation} 
\label{eq:tangent:identity}
\filtderiv{k;\parvec}[\md{k;\parvec}] = \pred{k;\parvec}\left[\left(\tstat{k} [\af{0:k}] - \pred{k;\parvec}[\tstat{k}[\af{0:k}]]\right) \md{k;\parvec}\right]\eqsp.
\end{equation}
It is worth noting that, in the context of this paper where $\qg{k}$ cannot be evaluated pointwise, one cannot expect to know the functional \eqref{eq:complete:gradient:log}, which involves the gradient of this quantity. 
In Section \ref{sec:simu:tangent:filter}, we illustrate that we can plug-in an estimate of this functionnal instead. The rationale motivating this algorithm relies on the following expression of the normalized loglikelihood:
$$
\frac{1}{n} \deriv \logllh{\parvec}(Y_{1:n}) = \frac{1}{n} \sum_{k = 1}^n \deriv \logllh{\parvec}\left(Y_k \mid Y_{1:k - 1}\right)\eqsp.$$
Moreover, under strong mixing assumptions, for all  $\parvec \in \parvec$, the extended process $\{ (X_n, Y_n, \pred{n}, \filtderiv{n}) \}_{n \geqslant 0}$ is an ergodic Markov chain and for all $\parvec \in \parvec$, the normalized score $\deriv \logllh{\parvec}(Y_{1:n})/n$  converges almost surely to a limiting quantity $\lambda(\parvec, \parvec_{\star})$ such that, under identifiability constraints, $\lambda(\parvec_{\star}, \parvec_{\star}) = 0$. 
A gradient ascent algorithm cannot be designed as the limiting function $\parvec \mapsto \lambda(\parvec, \parvec_{\star})$ is not available explicitly. 
However, solving the equation $\lambda(\parvec_{\star}, \parvec_{\star}) = 0$ may be cast into the framework of \emph{stochastic approximation} to produce parameter estimates using the \emph{Robbins-Monro algorithm}
\begin{equation}
\label{eq:par:update}
\parvec_{k} = \parvec_{k - 1} + \gamma_{k} \zeta_{k}\eqsp, \quad n\geqslant 0\eqsp, 
\end{equation}
where $\zeta_{k}$ is a noisy observation of $\lambda(\parvec_{k - 1}, \parvec_{\star})$, equal to \eqref{eq:online:gradient}.  
In the case of a finite state space $\set{X}$ the algorithm was studied in~\cite{legland1997recursive}, which also provides assumptions under which the sequence $\{\parvec_{n}\}_{n\geqslant 0}$ converges towards the parameter $\parvec_{\star}$ (see also \cite{tadic2010analyticity} for refinements). 

\subsection{Approximation of the filtering distribution}
\label{sec:filtering}
Let $(\epart{0}{\ell})_{\ell = 1}^\N$ be independent and identically distributed according to an instrumental proposal density $\rho_0$ on $\mathbb{R}^d$ and define the importance weights $\ewght{0}{\ell} \eqdef \Xinit(\epart{0}{\ell})/\XinitIS{0}(\epart{0}{\ell})$, where $\chi$ is the density of the distribution of $X_0$ as defined in Section~\ref{sec:model}. 
For any bounded and measurable function  $h$ defined on $\mathbb{R}^d$, the importance sampling estimator defined as 
\begin{align*}
\post{0}^\N[h]&\eqdef \sumwght{0}^{-1}\sum_{\ell=1}^\N \ewght{0}{\ell}h(\epart{0}{\ell})\eqsp,\quad\mbox{where}\quad\sumwght{0}\eqdef \sum_{\ell=1}^N \ewght{0}{\ell}\eqsp.
\end{align*}
is a consistent estimator of $\post{0}[f]$. 
Then, for all $k\geqslant 1$, once the observation $Y_k$ is available, \textit{particle filtering} transforms the weighted particle sample $\{(\ewght{k-1}{\ell},\epart{k-1}{\ell})\}_{\ell=1}^{\N}$ into a new weighted particle sample approximating $\post{k}$.
 This update step is carried through in two steps, \emph{selection} and \emph{mutation},  using sequential importance sampling and resampling steps.   
 New indices and particles $\{ (I_k^{\ell}, \epart{k}{\ell},\zeta_k^{\ell}) \}_{\ell = 1}^\N$ are simulated independently from the instrumental distribution with density on $\{1, \dots, \N\} \times \mathbb{R}^d\times \marginalset$:
$$
\instrpostaux{k}(\ell,x,z) \propto \ewght{k-1}{\ell} 
\kissforward{k-1}{k-1}(\epart{k-1}{\ell},x) \kernelmarg_{k}(\epart{k-1}{\ell},x;z) \eqsp,
$$
where 
$\kissforward{k-1}{k-1}$ is a Markovian transition density.  In practice, this step is performed as follows:
\begin{enumerate}
\item Sample $I_k^{\ell}$ in $\{1,\ldots,N\}$ with probabilities proportional to $\{\ewght{k-1}{j}
\}_{1\leqslant j\leqslant \N}$.
\item Sample $\epart{k}{\ell}$ with distribution $\kissforward{k-1}{k-1}(\epart{k-1}{I_k^{\ell}},\cdot)$ and sample $\zeta_k^{\ell}$ with distribution $\kernelmarg_{k}(\epart{k-1}{I_k^{\ell}},\epart{k}{\ell};\cdot)$.
\end{enumerate}
For any  $\ell \in\{1, \dots, \N\}$, $\epart{k}{\ell}$ is associated with the  importance weight defined by:
\begin{equation}
\label{eq:weight-update-filtering}
    \ewght{k}{\ell} \eqdef \frac{\hatqg{k-1}\langle\zeta_k^{\ell}\rangle(\epart{k-1}{I_k^{\ell}},\epart{k}{\ell}) }{ \kissforward{k-1}{k-1}(\epart{k-1}{I_k^{\ell}},\epart{k}{\ell})}
\end{equation}
to produce the following approximation of $\post{k}[f]$:
\begin{align*}
\post{k}^\N[f]&\eqdef \sumwght{k}^{-1}\sum_{\ell=1}^N \ewght{k}{\ell}f(\epart{k}{\ell})\eqsp,\quad\mbox{where}\quad\sumwght{k}\eqdef \sum_{\ell=1}^\N \ewght{k}{\ell}\eqsp.
\end{align*}
The choice of the proposal distribution $\kissforward{k-1}{k-1}$ is a pivotal tuning step to obtain efficient estimations of the filtering distributions.  This point will be discussed in each example considered in Section~\ref{sec:application}.

\subsection{Approximation of the backward statistics}
\label{sec:smoothing}

Approximation of the backward statistics, as defined in \eqref{eq:T:stat}, are computed recursively, for each simulated particle. 
The computations starts with initializing a set $\tstat[1]{0} = 0,\dots \tstat[N]{0} = 0$, corresponding to the values of $\tstat{0}(\epart{0}{1}), \dots, \tstat{0}(\epart{0}{N})$.  Then, using \eqref{eq:property:backward}, for each $k\geqslant 0$, $1\leqslant i\leqslant \N$, the approximated statistics are updated with:
\begin{equation}
\label{eq:update:paris:marginal}
\tstat[i]{k+1}  = \frac{1}{\K} \sum_{j=1}^{\K} \left( \tstat[\bi{k+1}{i}{j}]{k} + \addf{k}\left(\epart{k}{\bi{k+1}{i}{j}}, \epart{k+1}{i}\right)\right)\eqsp,
\end{equation}
where $\K\geqslant 1$ is a sample size which  is typically small compared to $\N$ and where $(\bi{k+1}{i}{j},\zeta_{k+1}^{(i,j)})$, $1\leqslant j\leqslant \K$, are i.i.d. in $\{1,\ldots,\N\}\times\marginalset$ with distribution 
$$
\overline \upsilon_k^i(\ell,z)\propto\ewght{k}{\ell}\hatqg{k}\langle z\rangle(\epart{k}{\ell},\epart{k+1}{i})\kernelmarg_{k}(\epart{k}{\ell},\epart{k+1}{i};z)\eqsp.
$$
As explained in \cite{gloaguen2021pseudo}, this recursive update requires to produce samples $\bi{k+1}{i}{j}$ distributed according to the marginal  distribution of $\overline \upsilon_k^i$, referred to as the backward kernel. In practice, this requires computationally intensive sampling procedures and the only proposed practical solution can be used in very restrictive situations.  
In \cite{gloaguen2018online}, the authors assumed that  almost surely, for all $x,x'$, $\hatqg{k}\langle \zeta\rangle(x, x')\geqslant 0$, and that, for all  $0\leqslant k\leqslant n$ and $0\leqslant i\leqslant N$, there exists an upper bound $\hkup^i_k$  such that
\begin{equation}
\mathrm{sup}_{\ell,\zeta}\;\hatqg{k}\langle \zeta\rangle(\epart{k}{\ell},\epart{k+1}{i})\leqslant \hkup^i_k\eqsp. \label{eq:AR:bound}
\end{equation}
Then,  if the positiveness assumption of  $\hatqg{k}\langle \zeta\rangle(x, x')$ is satistified, the sampling from the distribution $\overline\upsilon_k$ is possible thanks to condition \eqref{eq:AR:bound}, as, for all $(i, z) \in \lbrace 1,\dots,N\rbrace\times\marginalset$, 
$$
\ewght{k}{\ell}\hatqg{k}\langle z \rangle(\epart{k}{\ell},\epart{k+1}{i})\kernelmarg_{k}(\epart{k}{\ell},\epart{k+1}{i};z) \leqslant \hkup_k \ewght
{k}{\ell} \kernelmarg_{k}(\epart{k}{\ell},\epart{k+1}{i};z)\eqsp. 
$$
Therefore, the following  accept-reject mechanism algorithm may be used to sample from $\overline \upsilon_k^i$.
\begin{enumerate}
\item A candidate $(J^\ast,\zeta^\ast)$ is sampled in $\{1,\ldots, \N\}\times\marginalset$ as follows: 
\begin{enumerate}
\item $J^\ast$ is sampled with probabilities proportional to  $(\ewght{k}{\ell})_{\ell=1}^{\N}$ ;
\item $\zeta^\ast$ is sampled independently with distribution $\kernelmarg_{k}(\epart{k}{J^\ast},\epart{k+1}{i};\zeta^\ast)$.
\end{enumerate}
\item  $(J^\ast,\zeta^\ast)$ is then accepted with probability $\hatqg{k}\langle \zeta^\ast\rangle(\epart{k}{J^\ast}, \epart{k+1}{i})/\hkup_k^i$ and, upon acceptance, $\bi{k+1}{i}{j} = J^*$.
\end{enumerate}
This algorithm is the only online SMC smoother proposed in the literature with  theoretical guarantees when no closed-form expressions of  the  transition  densities and the conditional likelihood of the observations are available,  assuming that the user  can only evaluate  approximations of these densities. 
This pseudo-marginal particle smoothing algorithm  requires that the backward sampling step  generates samples exactly according to $\overline \upsilon_k^i$.  
However, it relies on the key assumptions of the positiveness of $\hatqg{k}\langle \zeta\rangle(x, x')$ and \eqref{eq:AR:bound} which are rather restrictive (especially the second one), and would not be satisfied in practice for a lot of problems (see for instance in Section~\ref{sec:simu:LV}).
 In Section \ref{sec:backwardis}, we propose an alternative to this step to obtain a computationally efficient pseudo-marginal smoother in a much wider range of applications for which such assumptions do not hold.
 
\subsubsection*{Approximations for recursive MLE}

In the case of recursive MLE, one needs to approximate the key quantity \eqref{eq:online:gradient}.
A particle filter, can be used to compute the following sequential Monte Carlo approximations:
$$
\pred{k}^N[\md{k;\parvec}] = \frac{1}{N}\sum_{\ell = 1}^N \md{k;\parvec}(\epart{n}{\ell})\eqsp,\eqsp \pred{k}^N[\deriv \md{k;\parvec}] = \frac{1}{N}\sum_{\ell = 1}^N \deriv \md{k;\parvec}(\epart{k}{\ell})\eqsp.
$$
In addition, the tangent filter can be approximated using a backward sampling procedure, based on the backward statistic associated with  the functional \eqref{eq:complete:gradient:log}:
\begin{equation} 
\label{eq:tangent:identity:part:linear}
\eta_{k;\parvec}^{N}[\md{k;\parvec}] = \frac{1}{\N}\sum_{\ell=1}^\N\tau_\ell^k \md{k;\parvec}
(\epart{\ell}{k}) - \left(\frac{1}{\N}\sum_{\ell=1}^\N\tau_\ell^n\right)\left(\frac{1}{\N}\sum_{\ell=1}^\N \md{k;\parvec}(\epart{\ell}{k})\right)\eqsp.
\end{equation}
Plugging these estimates in equation \eqref{eq:online:gradient} allows to perform the online recursive algorithm.

\section{Pseudo-marginal backward importance sampling}
\label{sec:backwardis}

\subsection{Positive estimates}
\label{sec:wald:trick}
In this section, we propose to use Wald’s identity for martingales to obtain an  estimator which is guaranteed to be positive.  This step is not required if $\hatqg{k}\langle z\rangle(x,x')$ is positive by construction but this is not necessarily true.
 Wald's trick was for instance applied in \cite{fearnhead2010random} to solve the filtering problem in the context of Poisson based estimators for partially observed diffusions. Our estimator is defined up to an unknown constant of proportionality, which is removed when the importance weights are normalized in equation \eqref{eq:update:paris:marginal:is}. 
This approach, rather than setting negative weights to 0, which would lead to a biased estimate, uses extra simulation to obtain positiveness. 
This is done while ensuring that the weights remain unbiased up to a common constant of proportionality.

\paragraph{Particle filtering weights.}  
For all $k\geqslant 0$, the Wald-based random weight particle filtering proceeds as follows.
\begin{enumerate}
\item For all $1\leqslant i\leqslant \N$, sample a new particle as described in Section~\ref{sec:filtering}.
\begin{enumerate}
\item Sample $I_k^{i}$ in $\{1,\ldots,\N\}$ with probabilities proportional to $\{\ewght{k-1}{j} 
\}_{1\leqslant j\leqslant \N}$.
\item Sample $\epart{k}{i}$ with distribution $\kissforward{k-1}{k-1}(\epart{k-1}{I_k^{i}},\cdot)$.
\end{enumerate}
\item For all $1\leqslant i\leqslant \N$, set  $\ewght{k}{i}= 0$.
\item While there exists $i_* \in\{1,\ldots,\N\}$ such that $\ewght{k}{i_*} \leqslant 0$, for all $1\leqslant i\leqslant \N$, sample $\zeta_k^{i}$ with distribution $\kernelmarg_{k}(\epart{k-1}{I_k^{i}},\epart{k}{i};\cdot)$ (i.e. compute an estimator of the transition density) and set 
$$
\ewght{k}{i}  = \ewght{k}{i} + \frac{\hatqg{k-1}\langle \zeta_k^i\rangle(\epart{k-1}{I_k^{i}},\epart{k}{i}) }{\kissforward{k-1}{k-1}(\epart{k-1}{I_k^{i}},\epart{k}{i})}\eqsp.
$$
\end{enumerate}
We aim at updating the backward statistics $\tstat[i]{k+1}$, $1\leqslant i \leqslant N$ using an importance sampling step as exact accept-reject sampling of the $\bi{k+1}{i}{j}$, $1\leqslant j \leqslant \K$,  with distribution $\overline \upsilon_k^i$ requires restrictive assumptions. Therefore, we introduce the following extension of Wald's trick importance sampling to the online smoothing setting of this paper.

\paragraph{Backward simulation weights.} For all $1\leqslant i\leqslant \N$, the backward importance sampling step proceeds then as follows.
\begin{enumerate}
\item For all $1\leqslant j\leqslant \K$,   sample $J_{k+1}^{(i,j)}$ in $\{1,\ldots, \N\}$ with probabilities proportional to  $(\omega^{i}_k)_{i=1}^{\N}$. 
\item For all $1\leqslant j\leqslant \K$, set  $\varpi_k^{(i,j)}= 0$.
\item While there exist $j_* \in\{1,\ldots,\K\}$ such that $\varpi_k^{(i,j)} \leqslant 0$, for all $1\leqslant j\leqslant \K$, sample $\zeta_k^{(i,j)}$  with distribution $\kernelmarg_{k}(\epart{k}{J_{k+1}^{(i,j)}}, \epart{k+1}{i};\cdot)$ and set 
$$
\varpi_k^{(i,j)} = \varpi_k^{(i,j)} +  \hatqg{k}\langle \zeta_k^{(i,j)} \rangle(\epart{k}{J_{k+1}^{(i,j)}}, \epart{k+1}{i})\eqsp.
$$
\end{enumerate}

\subsection{AR-free online smoothing}
Without any additional assumption, the statistics are then updated recursively as follows: for all $1\leqslant i\leqslant \N$,
\begin{equation}
\label{eq:update:paris:marginal:is}
\tstat[i]{k+1} =\sum_{j=1}^{\K}\frac{\varpi_k^{(i,j)}}{\mathcal{W}_k^i}\left( \tstat[\bi{k+1}{i}{j}]{k} + \addf{k}\left(\epart{k}{\bi{k+1}{i}{j}}, \epart{k+1}{i}\right)\right)\eqsp,
\end{equation}
where $\varpi_k^{(i,j)}$, $1\leqslant j \leqslant \K$ are computed using the pseudo-marginal smoothing technique combined with Wald's identity and $\mathcal{W}_k^i = \sum_{j=1}^{\K} \varpi_k^{(i,j)}$.
Then, the estimator of the conditional expectation of the additive functional $\post{0:n\mid n}^{} [\af{0:n}]$ is set as
\[
\post{0:n\mid n}^{\N,\textrm{IS}} [\af{0:n}] \eqdef \sum_{i=1}^{\N}\frac{\ewght{n}{i}}{\sumwght{n}}\tstat[i]{n}\eqsp.
\]

\section{Application to smoothing expectations and score estimation}
\label{sec:application}

\subsection{Recurrent neural networks}
\label{sec:simu:RNN}

Recurrent Neural Networks (RNNs) were first introduced in \cite{Mozer1989AFB} to model time series using a hidden context state. Such deep learning models are appealing to describe short time dependencies, and RNN extensions \cite{Hochreiter1997LongSM, Cho2014LearningPR} are since then widely used in practice, see for instance \cite{mikolov2010recurrent, sutskever2011generating, sutskever2014sequence}. 
 In this section, we propose a general state space model based on a vanilla RNN architecture, as follows.  
 The hidden state is initialized as $X_0 \sim \mathcal{N}(0,\Sigma)$ and for all $k\geqslant 1$,
$$
X_k = \tanh(W_{1} Y_{k-1} + W_{2} X_{k-1} + b + \eta_k)\quad\mathrm{and} \quad Y_k = W_{3} X_{k}  + c + \varepsilon_k\eqsp,
$$
where $W_{1}$, $W_{2}$, $W_3$ and $b$ and $c$ are the weight matrices and bias,  $\Sigma$ is an unknown covariance matrix and $(\eta_k)_{k\geqslant 1}$ and $(\varepsilon_k)_{k\geqslant 1}$ are independent Gaussian random variables with covariance matrices $Q$ and $R$.
In this experiment,  we show that the proposed algorithm can be used in this setting which does not fit the usual assumptions of hidden Markov models. 
While we here focus on a simple vanilla one-layer recurrent network, such general state space model could be extended to multi-layer RNN architectures by considering noisy state dynamics in each hidden layer, and to RNN variants such as Long Short Term Memory  \cite{Hochreiter1997LongSM} and Gated Recurrent Unit (GRU) \cite{Cho2014LearningPR}.
 
To generate a synthetic sequence of states and observations $(X_{0:n},Y_{0:n})$ from this stochastic RNN, we considered diagonal covariance matrices $\Sigma$, $Q$ and $R$, with the same variance along all dimensions equal to 0.1. 
 To obtain weights and biases values corresponding to realistic data, we  trained a deterministic one-layer RNN on 20,000 samples of a weather time-series dataset available online\footnote{https://www.bgc-jena.mpg.de/wetter/}. 
In such setting, the observations $Y_{0:n}$ consist in a sequence of 4D vectors (originally temperature, air pressure, air humidity, air density).
Following the classical RNN framework,  the sequence of hidden states $X_{0:n}$ is usually made of higher-dimensional vectors; the experiments were performed for two RNNs of respective hidden dimension 32 and 64. 
After this training part, we sampled $(X_{0:n}, Y_{0:n})$ according to the model. 
For each RNN,  a single sequence of  hidden states and observations was simulated with a total length of 200 time steps.
 
 From this general state space model based on a stochastic RNN, in the context of the \textit{state estimation} problem, we are interested in using particle smoothing algorithms to estimate two smoothing expectations, respectively $\mathbb{E}[X_0|Y_{0:n}]$, and $\mathbb{E}[\sum_{k=0}^{n} X_k|Y_{0:n}]$. 
 In our context of online estimation, the evaluation was made for $n=49$ (sequence truncated at 50 observations), $n=99$ (sequence truncated at 100 observations), and $n=199$ (full sequence). 
  The Monte Carlo  estimate of these quantities is referred to with the hat symbol: $\mathbb{\widehat{E}}[.]$. 

In the following, the performances of our algorithm was compared to the classical Poor Man's smoother. 
The  Poor Man's smoother (also known as the path-space smoother) estimates the joint smoothing distribution using the ancestral lines of each particle at time $n$; see for instance \cite{douc2014nonlinear} for discussions on the path degeneracy issue.
For the backward IS smoother, we use $N=1000$ particles for the bootstrap filter and $\K=32$ backward samples (see Section \ref{sec:simu:SINE} and Figure~\ref{fig:sine:timeandbias:N:vary} for the choice of $\K$ from the number of particles $N$). 
For the Poor Man's smoother $N=3000$ particles were used, which yields a similar computational cost than the backward IS smoother. 
One interesting aspect of applying the backward IS smoother (BIS) on neural network architectures is the parallelization abilities of such algorithm: the loop over the backward samples in the backward step of the BIS is easily parallelizable, while the Poor Man's smoother requires to store the full past trajectories of the particles. 

\begin{figure}[h]
\begin{center}
    \includegraphics[width=\textwidth, trim = 1cm 1cm 1cm 1cm, clip]{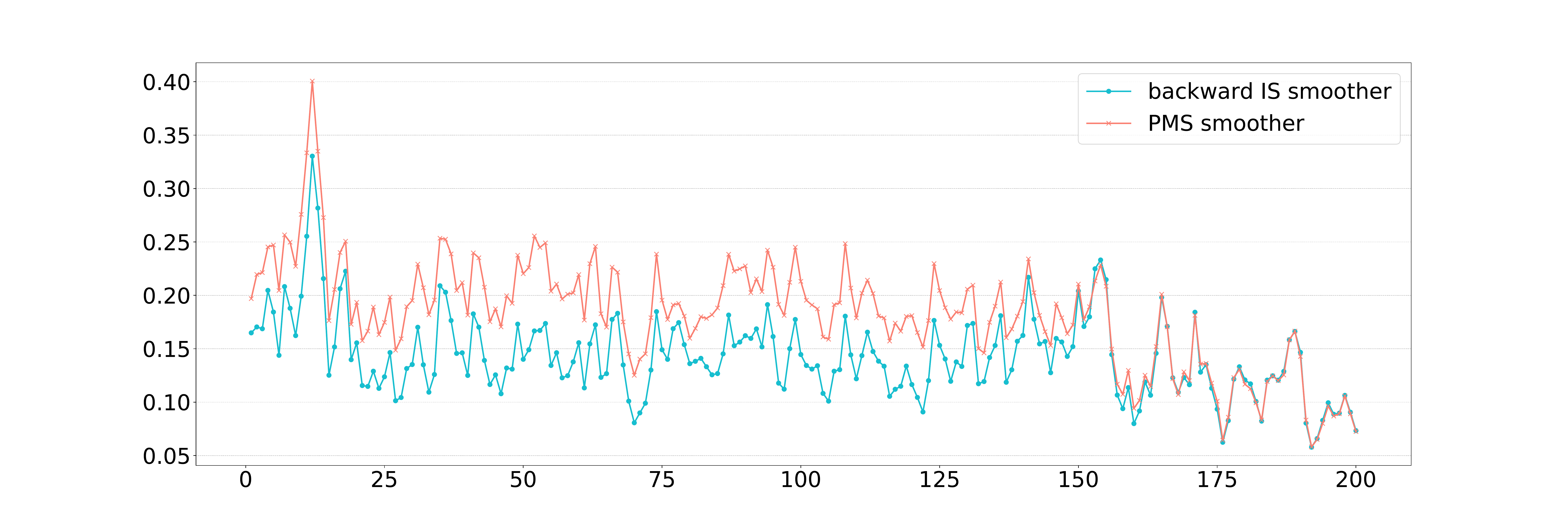}
\end{center}
    \caption{Plot of the empirical estimate of $\mathbb{E}[\|X_k - \mathbb{\widehat{E}}[X_k|Y_{0:199}]\|^2]$ for $k \in \{0,...,199\}$ for 100 runs of the Backward IS and Poor Man smoothers.}
    \label{fig:RNN:mseperXk}
\end{figure}

Table~\ref{table:RNN_exp} displays the result when performing 100 runs for each smoothing algorithm. 
The performance metric is the classical mean squared error (MSE), which is approximated with the empirical mean over the 100 runs.
The backward IS smoother outperforms the Poor Man's smoother when estimating both quantities. 
This is also illustrated in Figure~\ref{fig:RNN:mseperXk}, displaying for the stochastic RNN of dimension $64$ the MSE over 100 runs of $\mathbb{\widehat{E}}[X_k|Y_{0:199}]$, for $k \in \{0,...,199\}$: the backward IS smoother has a significantly  smaller MSE than the Poor Man's smoother for all observations that are recorded far in the past (in our example, for all $k \leqslant 150$).  
Moreover, the table also shows that while the backward IS's MSE tends to stay stable for all given $n$ (49,99, and 199), as expected the Poor Man's estimation is less accurate for a longer sequence of observations, with a MSE increasing as $n$ increases.



\begin{table}
\caption{Empirical estimation of $\mathbb{E}[\|X_0 - \mathbb{\widehat{E}}[X_0|Y_{0:n}]\|^2]$ and $\mathbb{E}[ \sum_{k=0}^{n}\|X_k - \mathbb{\widehat{E}}[X_k|Y_{0:n}]\|^2/(n+1)]$ for the Poor man's smoother (PMS) and the Backward IS smoother, for states $X_{0:n}$ and observations $Y_{0:n}$ generated with stochastic RNNs of dimension 32 and 64. We consider an online estimation of a sequence of 200 observations, truncated at timestep $n=49$ (50 timesteps), $n=99$ (100 timesteps) and the full sequence ($n=199$).}
\label{table:RNN_exp}
\centering 
\begin{tabular}{c|cc|cc}
 & \multicolumn{2}{c}{RNN dim = 32} & \multicolumn{2}{c}{RNN dim = 64} \\\toprule
\hline
 &  PMS & Backward IS &  PMS & Backward IS \\
\hline
$\|X_0 - \mathbb{\widehat{E}}[X_0|Y_{0:199}]\|^2$ & 0.2147 & 0.1997 & 0.1969 & 0.1649\\
$ \sum_{k=0}^{199} \|X_k - \mathbb{\widehat{E}}[X_k|Y_{0:199}]\|^2/200$ & 0.2551 & 0.2056  & 0.1822 & 0.1427 \\
\hline
$\|X_0 - \mathbb{\widehat{E}}[X_0|Y_{0:99}]\|^2$ & 0.2135 & 0.1997 & 0.1914 & 0.1649 \\
$ \sum_{k=0}^{99} \|X_k - \mathbb{\widehat{E}}[X_k|Y_{0:99}]\|^2/100$ & 0.2531 & 0.2189 & 0.1775 & 0.1519\\
\hline
$\|X_0 - \mathbb{\widehat{E}}[X_0|Y_{0:49}]\|^2$ & 0.2018 & 0.1997 & 0.1707 & 0.1650\\
$\sum_{k=0}^{49} \|X_k - \mathbb{\widehat{E}}[X_k|Y_{0:49}]\|^2/50$ & 0.2307 & 0.2147 & 0.1633 & 0.1589\\
\hline
\bottomrule
\end{tabular}
\end{table}

\subsection{One dimensional diffusion processe: the Sine model}
\label{sec:simu:SINE}
This section investigates the performance of the proposed algorithm to compute  expectations under the smoothing distributions in a context where  alternatives are available for comparison. Consider the Sine model where $(X_t)_{t\geqslant 0}$ is assumed to be a weak solution to
$$
\rmd X_t = \sin(X_t-\theta)\rmd t + \rmd W_t\eqsp,\quad X_0 = x_0\eqsp.
$$
This simple model has no explicit transition density, however, a General Poisson estimator which satisfies \eqref{eq:AR:bound} can be computed by simulating Brownian bridges, (see \cite{beskos2006exact}). 
Therefore, the backward importance sampling technique proposed in this paper can be compared to the usual acceptance-rejection algorithm described in Section~\ref{sec:smoothing}. 
For this simple comparison, observations are received at evenly spaced times $t_0=0,\ldots, t_{10} = 5$ from the model
\begin{equation}
\label{eq:obs:model:SINE}
Y_k=X_{t_k}+\varepsilon_k,\eqsp 0\leqslant k \leqslant n = 10\eqsp,
\end{equation}
where $(\varepsilon_k)_{0 \leqslant k\leqslant 10}$ are i.i.d. Gaussian random variables with mean $0$ and variance $1$. In this experiment $\theta = \pi/4$. 
The proposal distribution $p_k$ for the particle filtering approximation is chosen as the following approximation of the optimal filter:
\begin{equation}
\label{eq:optimal:filter}
p_{k}(x_{k},x_{k+1})\!\propto\! q^{\mathsf{Eul}}_{k+1}(x_{k},x_{k+1})g_{k+1}(x_{k+1},Y_{k+1})\eqsp,
\end{equation}
where $q^{\mathsf{Eul}}_{k+1}$ is the probability density function of Gaussian distibution with mean $\Delta \sin(x_k-\theta)$ and variance $\Delta$ where $\Delta = 1/2$, i.e. the Euler approximation of the Sine SDE, and $g_k$ is the probability density function of the law of $Y_k$ given $X_{t_k}$ i.e. of a Gaussian random variable with mean $X_{t_k}$ and variance 1. 
As the observation model is linear and Gaussian, the proposal distribution is therefore Gaussian with explicit mean and variance. 

\begin{figure}[h]
\begin{center}
\includegraphics[scale = .4]{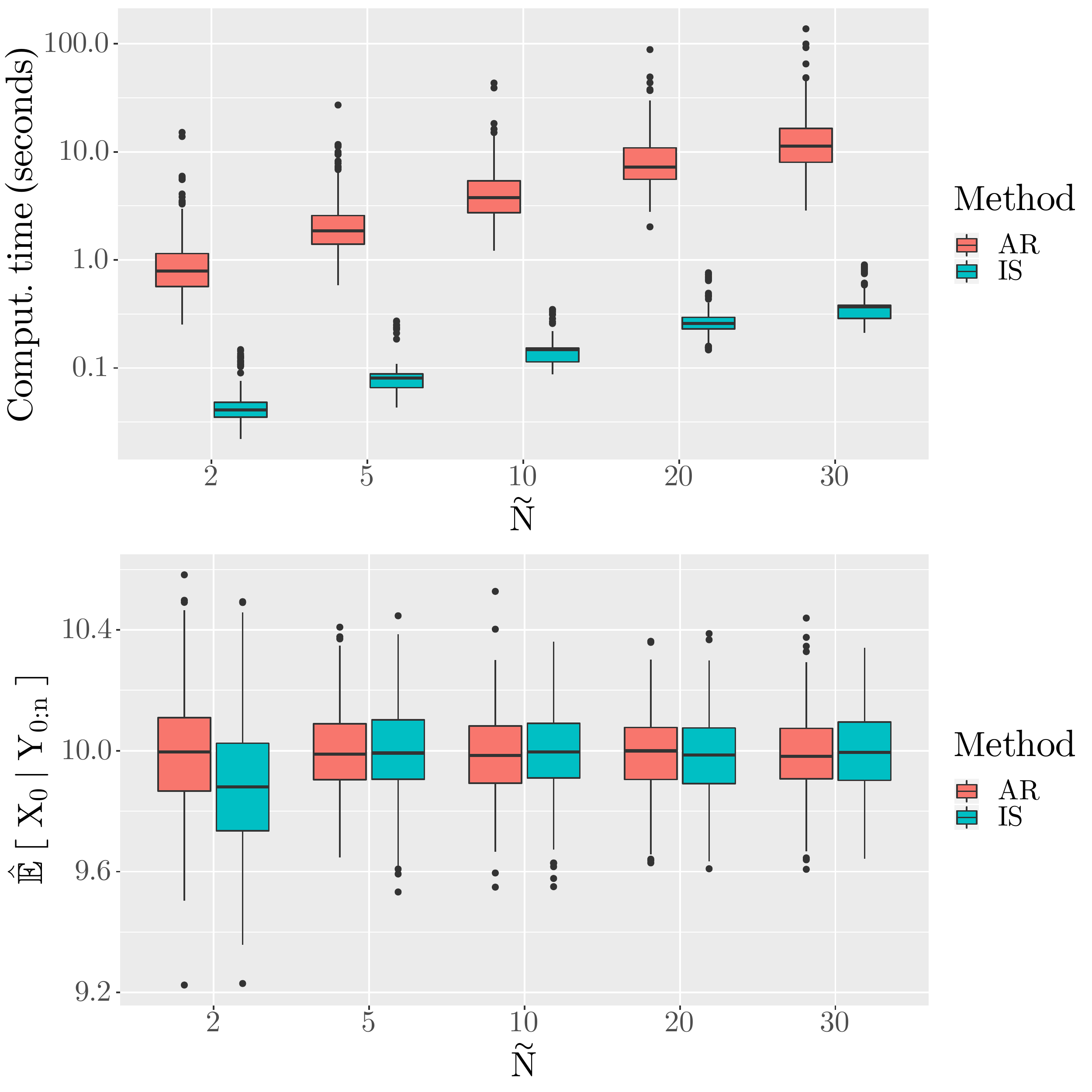}
\end{center}
\caption{Computational complexity and estimation of a posterior mean as a function of the number of backward samples. Results are shown for the state of the art acceptance-rejection algorithm and the proposed backward importance sampling technique.}
\label{fig:sine:timeandbias}
\end{figure}

In this first experiment, particles are used to solve the state estimation problem for the first observation i.e. to compute an estimate of $ \pE[ X_{0} | Y_{0:n}]$.
Figure~\ref{fig:sine:timeandbias} displays the computational complexity and the estimation of the posterior mean with the acceptance-rejection algorithm and the proposed backward sampling technique as a function of $\K$. 
In this setting, $N=100$, and each unbiased estimate of $\hat{q}$ is computed using 30 Monte Carlo replicates.

For $\K = 2$ (which is the recommended value for the PaRIS algorithm, see \cite{olsson2017efficient}), our estimate shows a bias, which is no surprise, as it is based on a biased normalized importance sampling step. However, this bias quickly vanishes for $\K \geqslant 10$. 
Interestingly, our method comes with a drastic (a factor 10) reduction of computational time. 
The vanishing of the bias might induce more backward sampling, but this remains much faster than the acceptance rejection method with $\K = 2$.

\begin{figure}[h]
\begin{center}
\includegraphics[scale = .4]{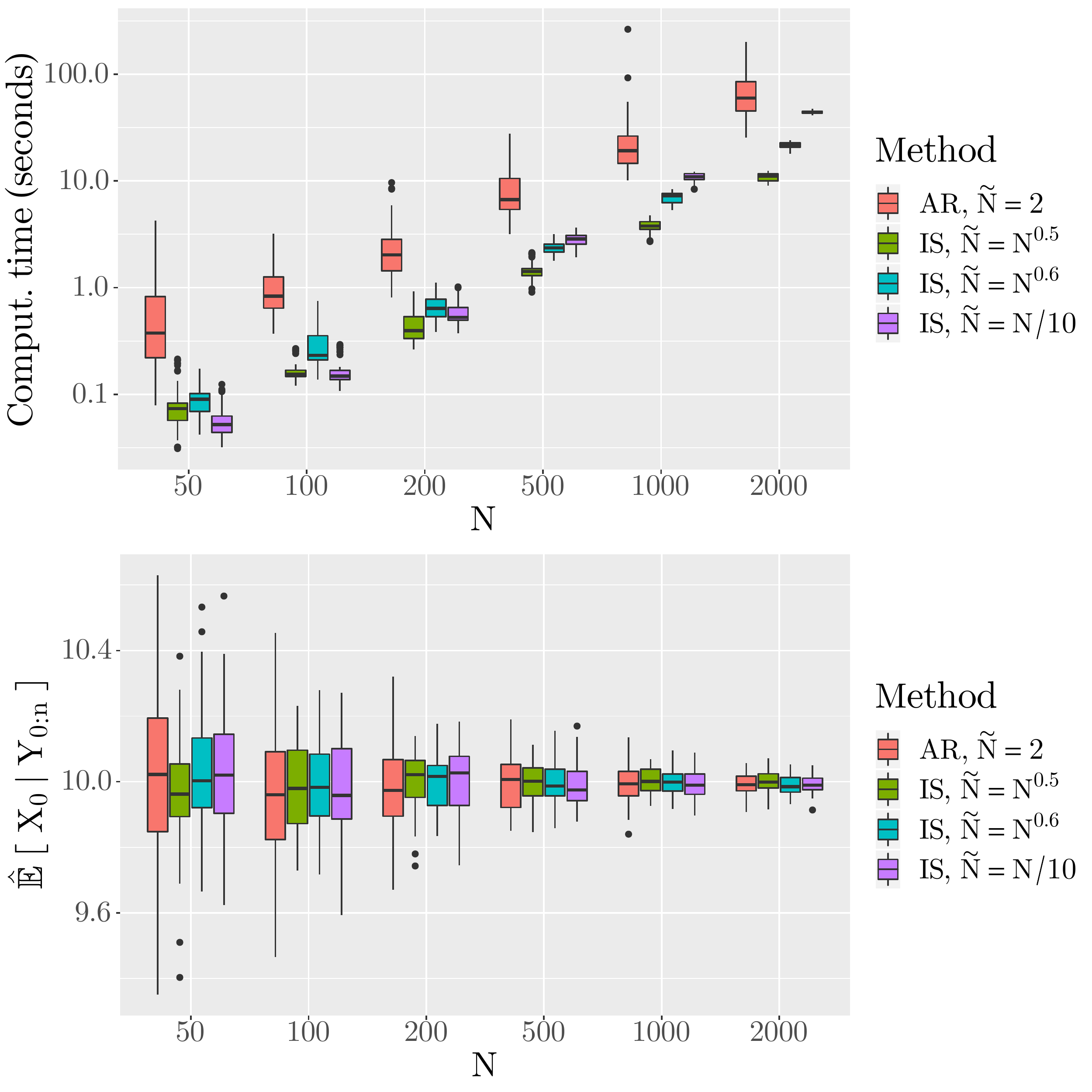}
\end{center}
\caption{Computational complexity and estimation of a posterior mean as a function of the number of particles. Results are shown for the state of the art acceptance-rejection algorithm and the proposed backward  importance sampling technique. The number of backward samples is set to 2 for the AR, and $N/10$ for the IS.}
\label{fig:sine:timeandbias:N:vary}
\end{figure}

Then, the same estimation was performed (on the same data set) for $\N$ varying from 50 to 2000.
In this context, $\K$ was set to 2 for the AR method. 
To have an empirical intuition of how $\K$ must vary with $\N$ for our algorithm, the backward importance sampling  is applied with $\K = \N^{0.5}, \N^{0.6}$ and $N / 10$ (as this last value was sufficient in the first experiment to avoid any bias). The results are shown in Figure~\ref{fig:sine:timeandbias:N:vary}. A  small bias might appear for $\N = 2000$ and $\K~=~45\eqsp(\approx 2000^{0.5})$, but no bias is visible for $\N^{0.6}$ and $\N /10$. 
As expected, the gain in time, compared to the state of the art algorithm, remains important (even if it decreases as $\K$ increases). 
It is worth noting that the variance of the computational time is greatly reduced compared to the AR technique.

\subsection{Recursive maximum likelihood estimation in the  Sine model}
\label{sec:simu:tangent:filter}
\begin{figure}
\centering
\includegraphics[width = 0.7\textwidth]{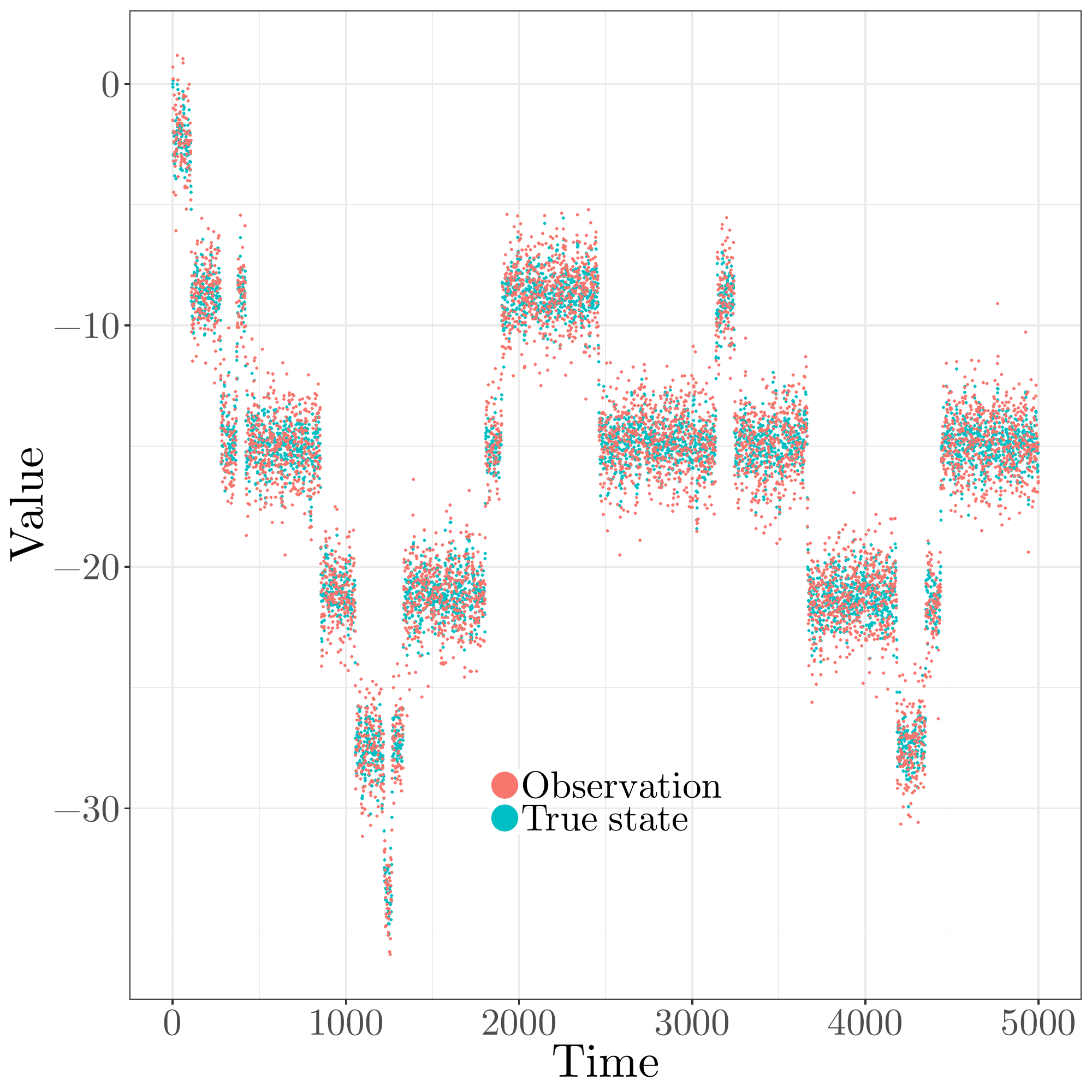}
\caption{\label{fig:data} Data set simulated according to the SINE process, observed with noise at discrete time steps.}
\end{figure}

Online recursive maximum likelihood using pseudo marginal SMC is illustrated for the same Sine model.
As mentionned, a GPE estimator of the transition density can be computed. 
Following the idea of this computation, it is possible to obtain an unbiased estimate of the gradient of the log-transition density and thus compute and unbiased estimate of the key quantity given in \eqref{eq:complete:gradient:log}. 
To the best of our knowledge, this estimator is new, and given in appendix \ref{sec:filter:SDE}.
Using the Exact algorithm of \cite{beskos2006retrospective} a data set of 5000 points (displayed in Figure~\ref{fig:data}), was simulated whith the true parameter $\parvec_* = \pi/4$. 
As in the previous section, particle smoothing was performed, using  the same particle filter, with $\N = 100$ particles and $\K = 10$ backward samples in our backward importance sampling procedure. 
In this setup, we explore three key features of our estimator.

\begin{figure}
\centering
\begin{tabular}{cc}
\includegraphics[width = 0.4\textwidth]{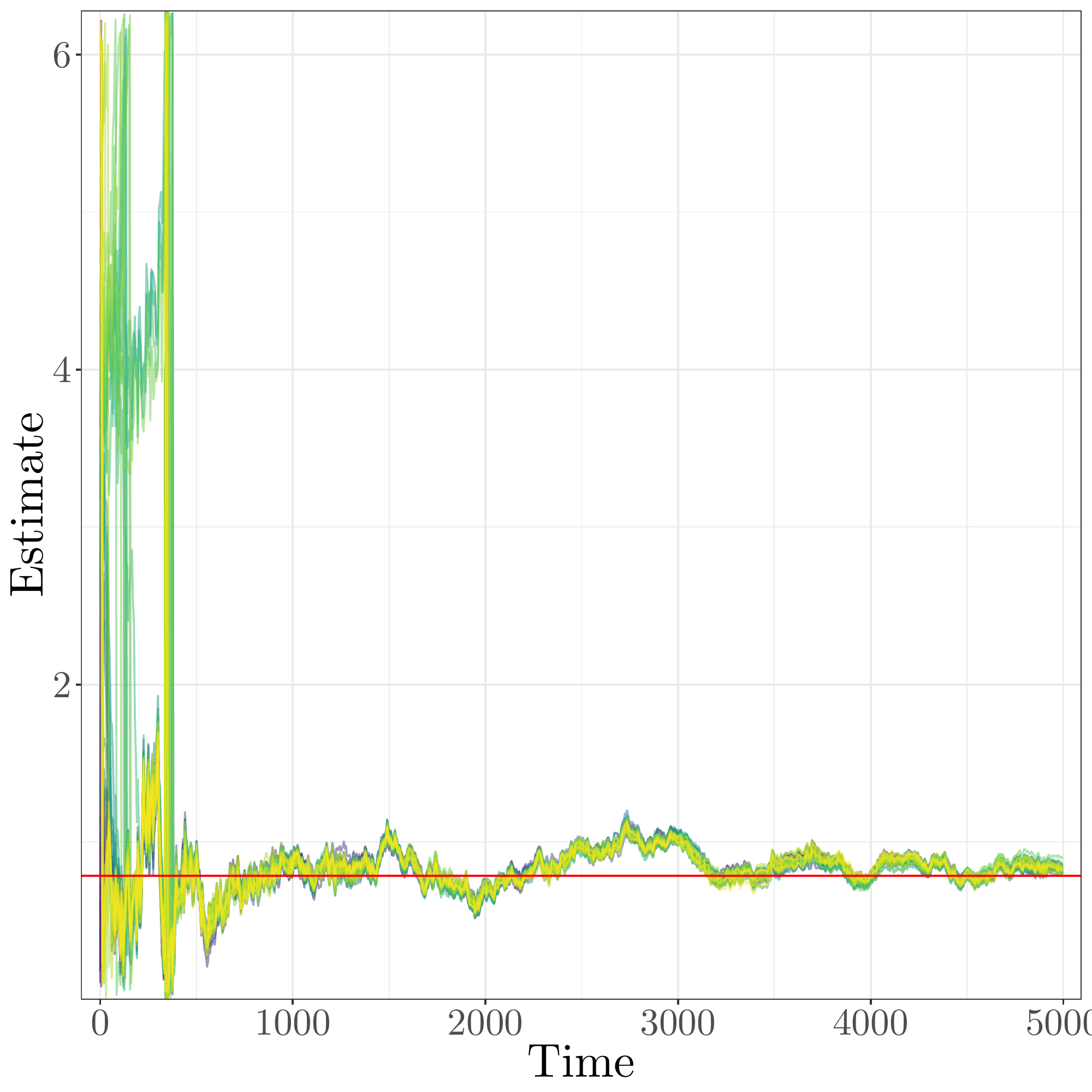}&
\includegraphics[width = 0.4\textwidth]{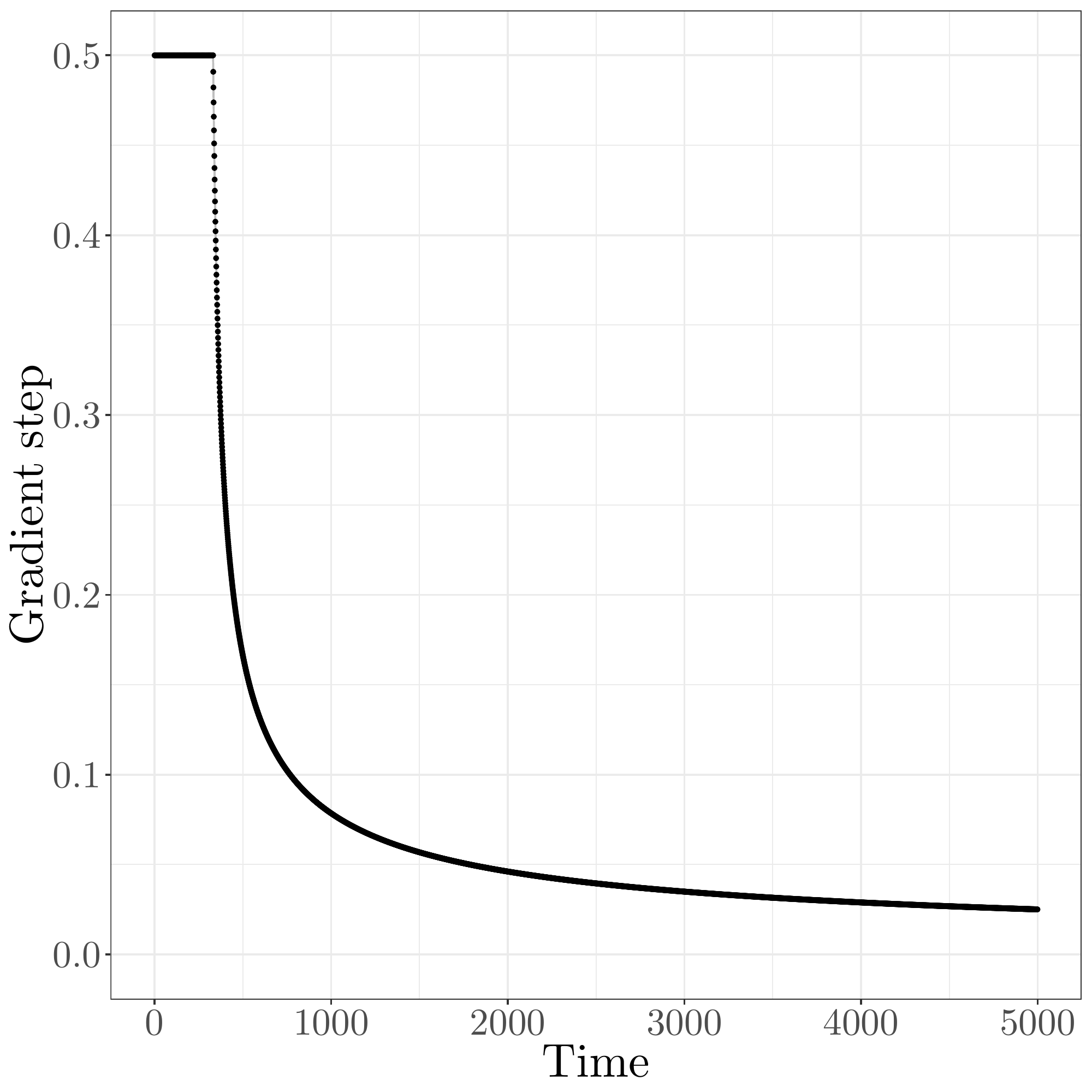}
\end{tabular} 
\caption{\label{fig:1obs:50start}(\textit{Left}) online estimation of $\parvec$ for the data set presented in Figure~\ref{fig:data}. The algorithm is performed from 50 starting points. (\textit{Right}) The gradient step sizes (defined in equation \eqref{eq:par:update}).}
\end{figure}

{\em Sensitivity to the starting point $\hat{\parvec}_0$.}
The inference procedure was performed on the same data set from 50 different starting points uniformly chosen in $(0,2\pi)$. 
The gradient step size $\gamma_k$ of equation \eqref{eq:par:update} was chosen constant (and equal to 0.5) for the first 300 time steps, and then decreasing with a rate proportional to $k^{-0.6}$. 
Results are given Figure~\ref{fig:1obs:50start}. There is no sensitivity to the starting point of the algorithm, and after a couple of hundred observations, the estimates all concentrate around the true value. 
As the gradient step size decreases, the estimates stay around the true value following autocorrelated patterns that are common to all trajectories.

\begin{figure}
\centering
\begin{tabular}{ccc}
\includegraphics[width = 0.32\textwidth]{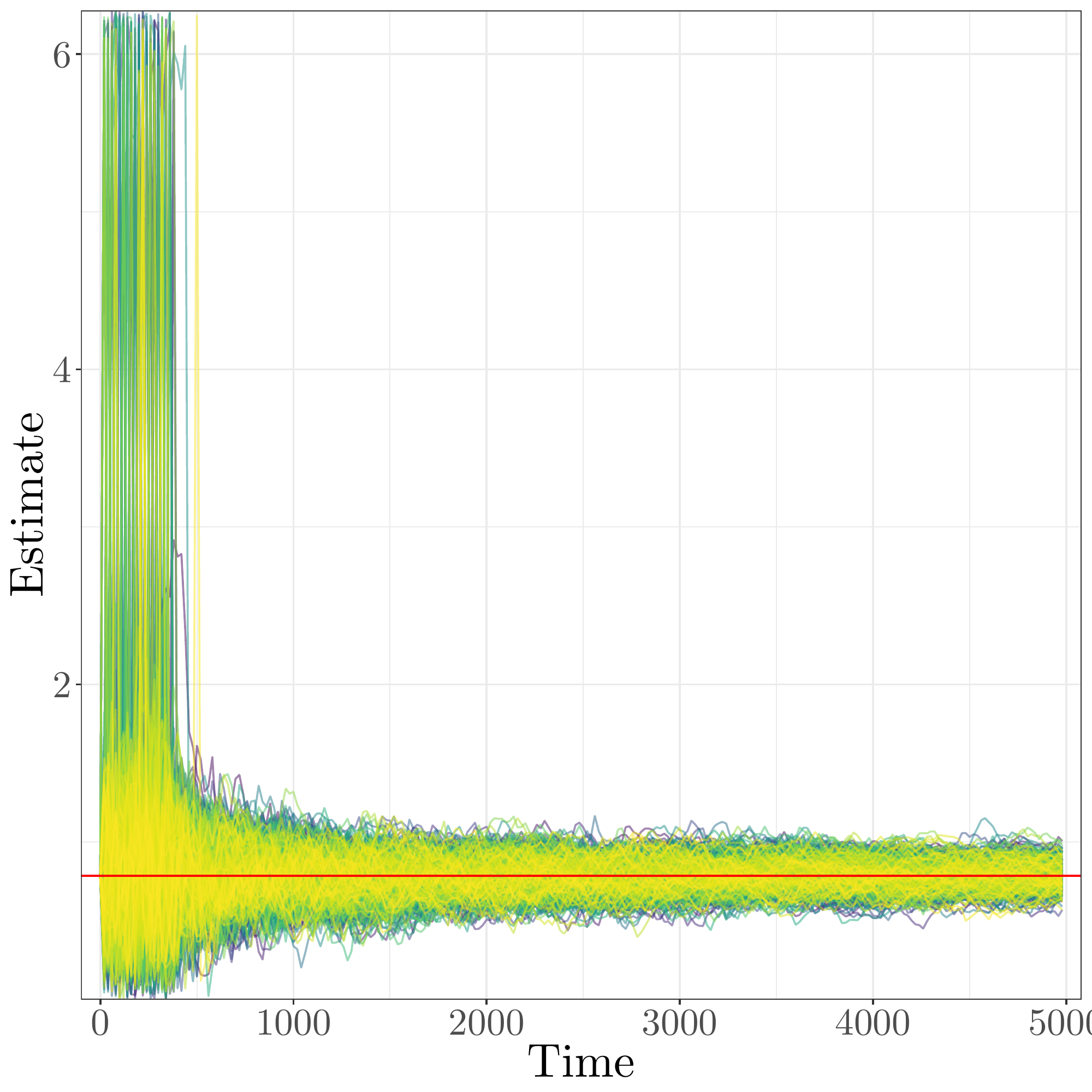}&
\includegraphics[width = 0.32\textwidth]{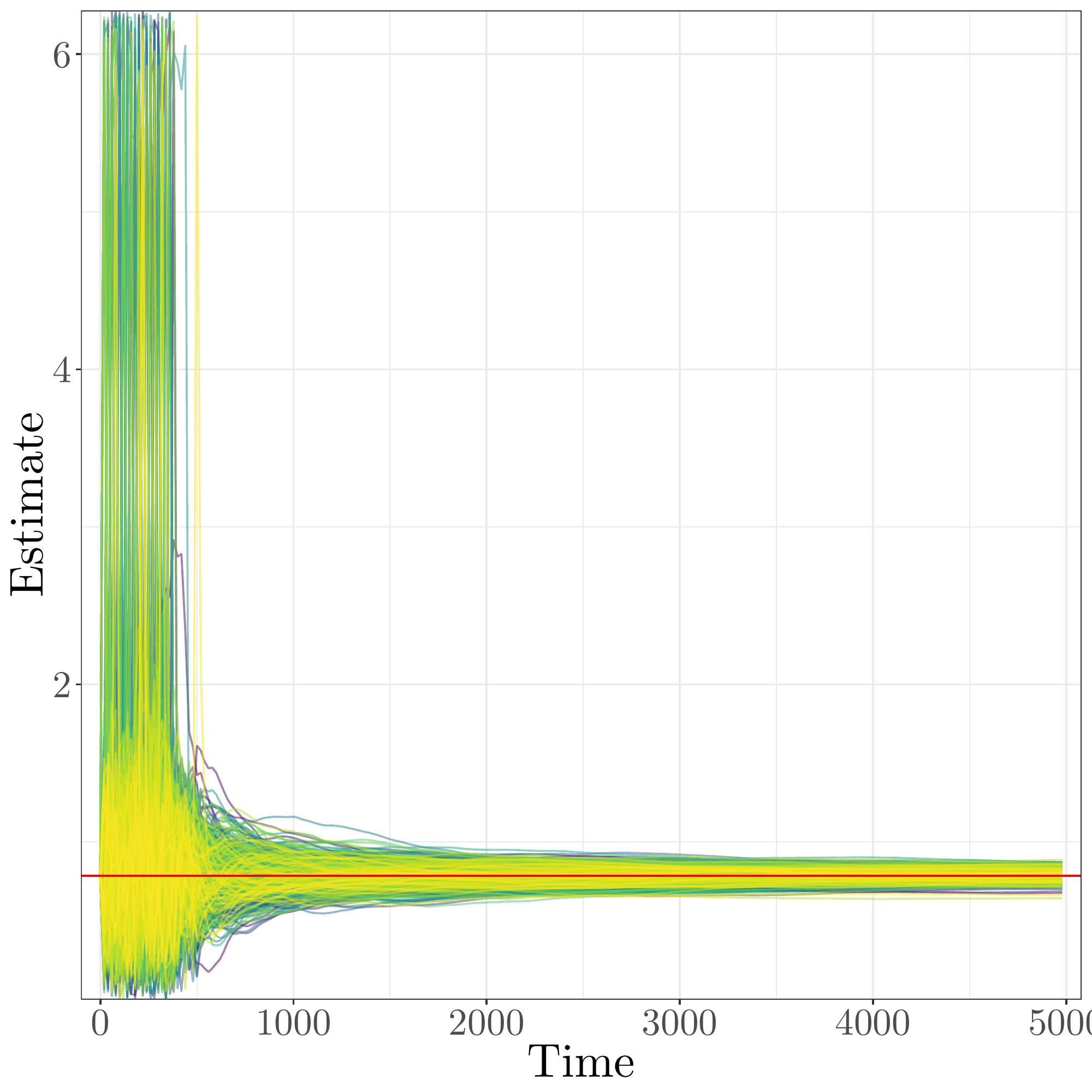}&
\includegraphics[width = 0.32\textwidth]{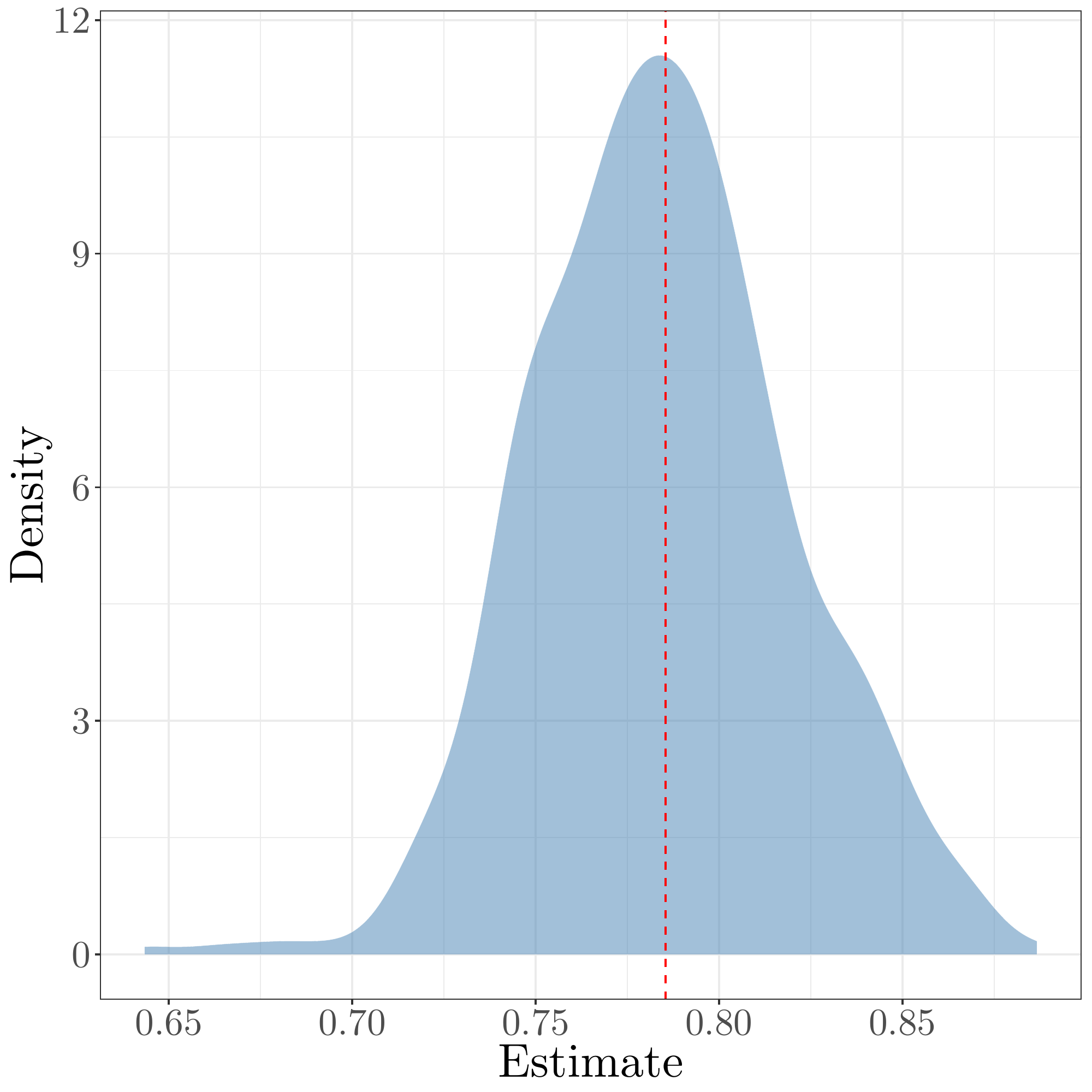}
\end{tabular}
\caption{\label{fig:50obs:1start}(\textit{Left}) online estimation of $\parvec$ for 50 different simulated data sets as presented in Figure \ref{fig:data}. The algorithm is performed from 1 starting point with the gradient step size shown in Figure \ref{fig:1obs:50start}. (\textit{Center}) Averaged estimator, where $\hat{\parvec}$ is averaged after a burning phase of 300 time steps. (\textit{Right}) Empirical distribution of $\hat{\parvec}$. The red line is the value of $\parvec^*$.}
\end{figure}

{\em Asymptotic normality.}
The inference procedure was performed on 50 different data sets simulated with the same $\theta_*$. 
The 50 estimates were obtained starting from the same starting point (fixed to $\theta_*$, as Figure \ref{fig:1obs:50start} shows no sensitivity to the starting point). Figure \ref{fig:50obs:1start} shows the results for the raw and the averaged estimates. 
 The averaged estimates $(\widetilde \parvec_k)_{k\geqslant 0}$ consist in averaging the values produced by the estimation procedure after a burning phase of $n_0$ time steps (here $n_0 =300$ time steps). 
 This procedure allows to obtain an estimator whose convergence rate does not depend on the step sizes chosen by the user, see \cite{polyak1992acceleration,kushner1997stochastic}. For all $0 \leqslant k\leqslant n_0$, $\widetilde \parvec_k = \widehat\parvec_k $ and for all $k>n_0$,
\[
\widetilde \parvec_k = \frac{1}{k-n_0}\sum_{j= n_0+1}^k\widehat\parvec_j\eqsp. 
\]
The estimated distribution of the final estimates seems to be Gaussian, centered around the true value. 
This conjecture, which would extend the asymptotic normality obtained in \cite{gloaguen2021pseudo} for the original pseudo-marginal PaRIS, should be proven in future works.
 
\begin{figure}
\centering
\begin{tabular}{cc}
\includegraphics[width = 0.4\textwidth]{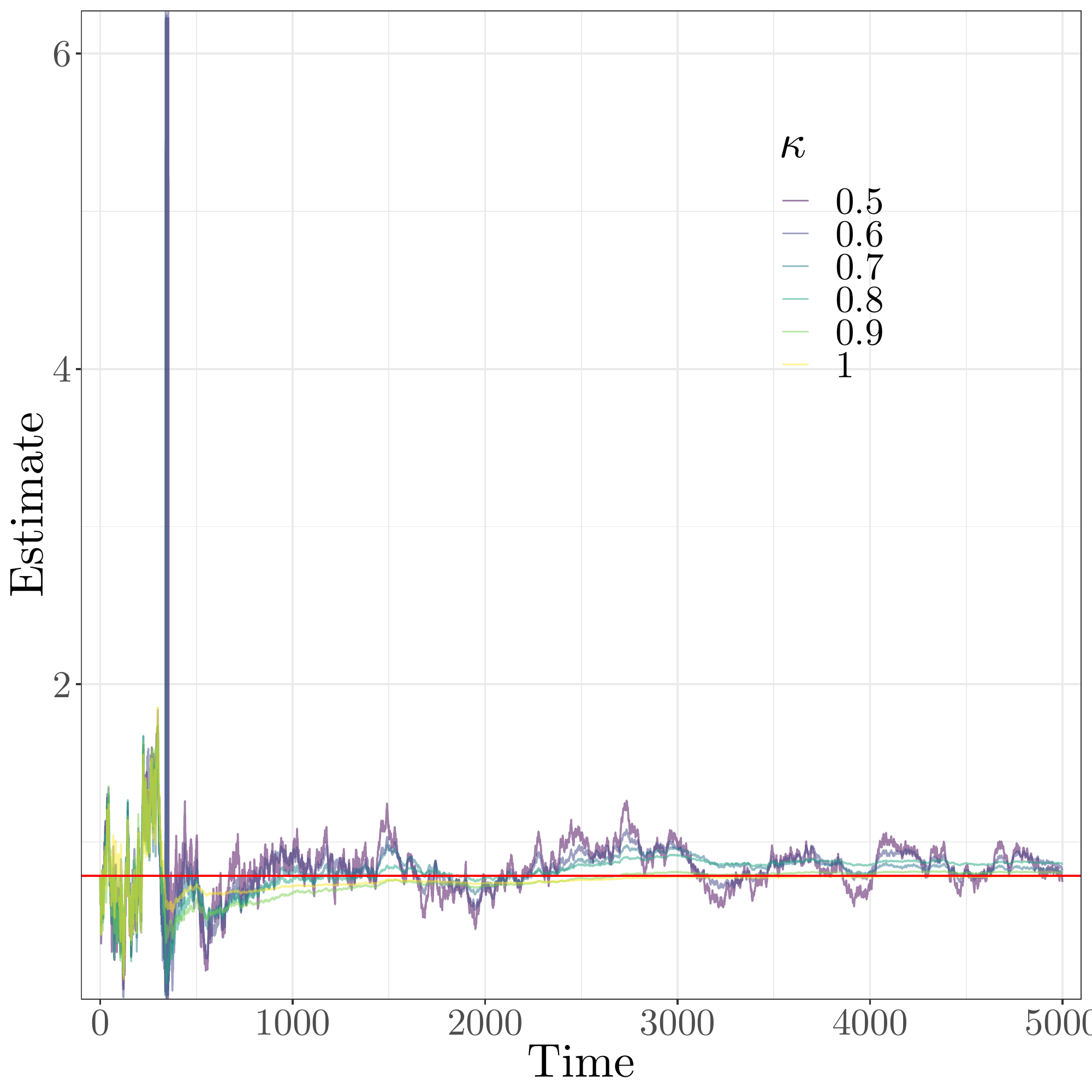}&
\includegraphics[width = 0.4\textwidth]{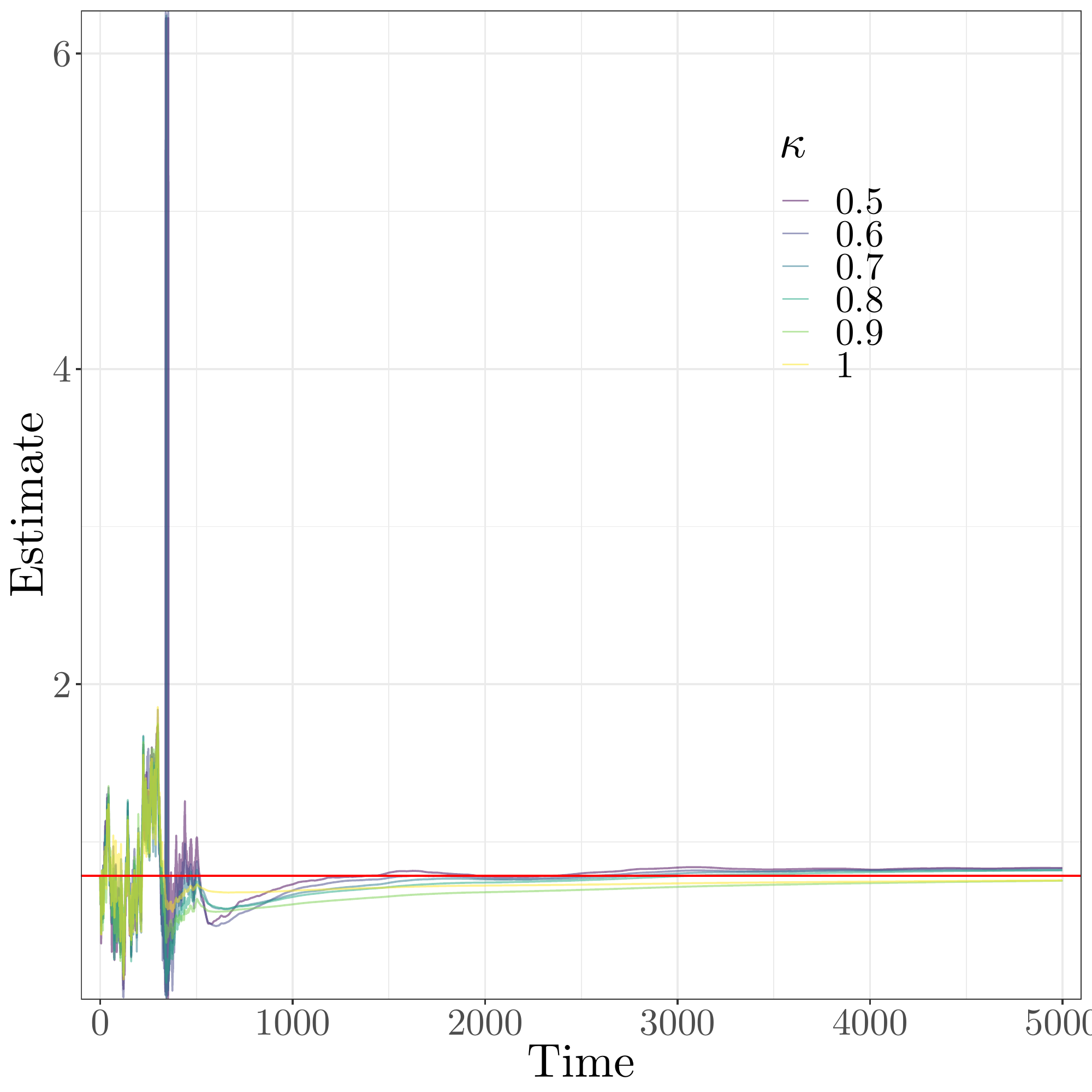}
\end{tabular}
\caption{\label{fig:1obs:1start:6Grads}(\textit{Left}) online estimation of $\theta$ for the data set presented in Figure \ref{fig:data}, with different decreasing rates values $\kappa$. (\textit{Right}) Averaged estimator, where $\hat{\parvec}$ is averaged after a burning phase of 300 time steps.}
\end{figure}

{\em Step size influence.} To illustrate the  influence of the gradient step sizes, different settings are considered. 
In each scenario, the sequence $(\gamma_k)_{k\geqslant 0}$ is given by
\begin{equation*}
\gamma_k = \gamma_0 \1_{\left\lbrace k \leq n_0 \right\rbrace} + \frac{\gamma_0}{(k - n_0)^{\kappa}}\1_{\left\lbrace k > n_0 \right\rbrace}\eqsp,
\end{equation*}
where $\gamma_0 = 0.5$. In this experiment $\kappa\in\{0.5,0.6,0.7,0.8,0.9,1\}$.
The results are shown in Figure~\ref{fig:1obs:1start:6Grads}. 
As expected, the raw estimator shows different rates of convergence depending on $\kappa$, whereas the averaged estimator has the same behavior in all cases. 


\begin{figure}
\begin{center}
\includegraphics[scale = 0.4]{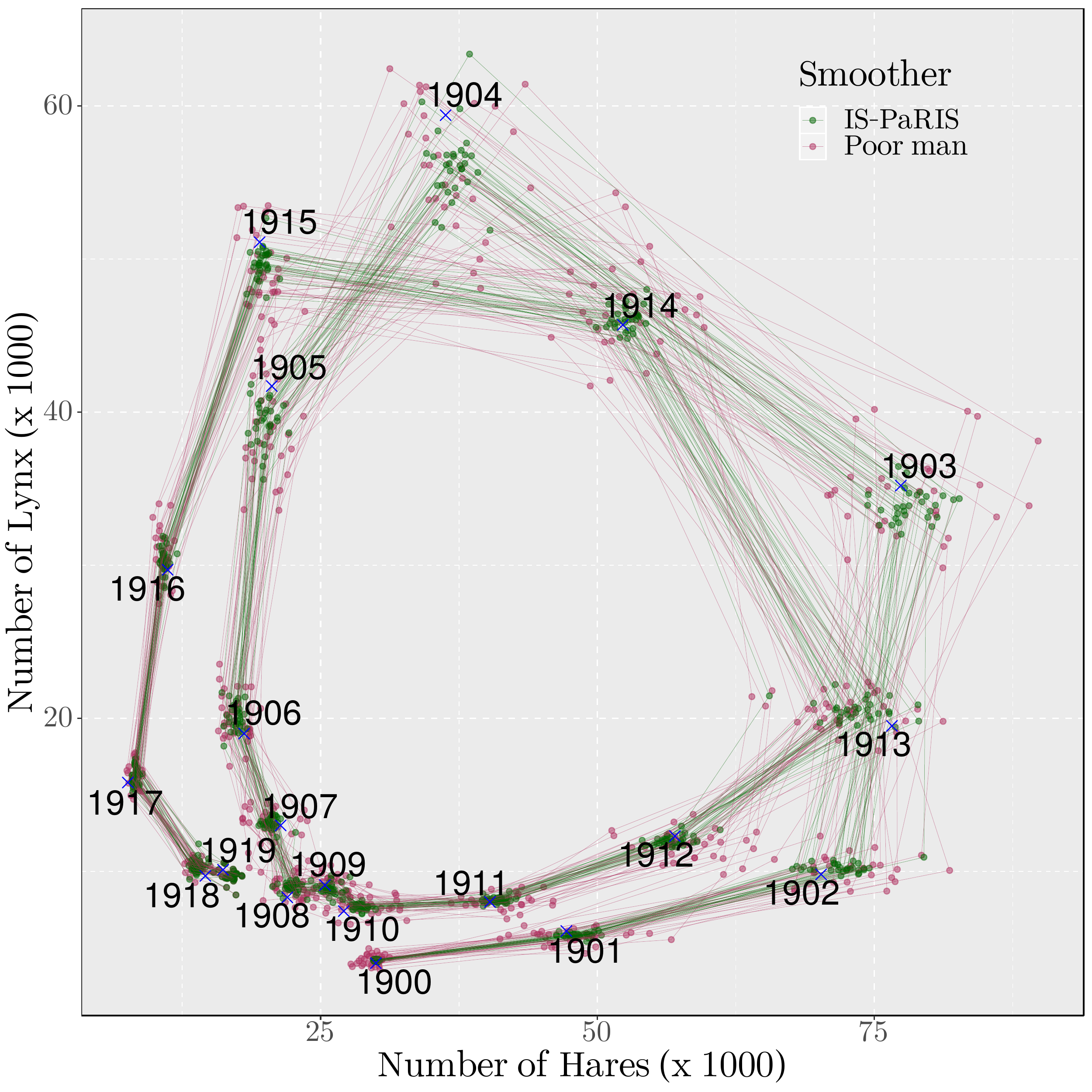}
\end{center}
\caption{\label{fig:LV:hares:lynx} Estimated Hares-Lynx abundances using the Hudson bay company data set. Both our IS-PaRIS smoother and the poor man smoother are performed to approximate the MLE and solve the tracking problem. Blue crosses show the observations.}
\end{figure}

\subsection{Multidimensional diffusion processes: Stochastic Lotka-Volterra model}
\label{sec:simu:LV}

This section sets the focus on a stochastic model describing in continuous time the population dynamics in a predator-prey system, as fully discussed in \cite{hening2018persistence}. 
The bivariate process $(X_t)_{t\geqslant 0}$ of predators and preys abundances is assumed to follow the stochastic  Lotka-Volterra model:
\begin{equation}
\label{eq:LV:SDE}
\rmd X_t = \alpha_\parvec(X_t) \rmd t + \begin{pmatrix}X_1(t) & 0 \\ 0 & X_2(t)\end{pmatrix} \Gamma \rmd \W_t\eqsp,\
\end{equation}
where $\W_t$ is a vector of independent standard Wiener processes, $\Gamma$ a $2\times 2$ matrix, and for $x = (x_1, x_2)^T:$
\[
\alpha_\parvec(x) = \begin{pmatrix} x_1( a_{10} - a_{11}x_1 - a_{12}x_2)\\  x_2(-a_{20} + a_{21}x_1 - a_{22}x_2) \end{pmatrix}\eqsp .
\]
In this context, the unknow parameter to be estimated is $\parvec = ( a_{10}, a_{11}, a_{12}, a_{20}, a_{21}, a_{22}, \Gamma)$. The observation model follows a widespread framework in ecology where the abundance of preys and predators are observed through some abundance index at discrete times $t_0, \dots, t_n$:
\begin{equation}
Y_{t_k} = \begin{pmatrix} \text{c}_1X_1(t_k)\mathrm{e}^{\epsilon^{(1)}_{t_k}} \\ \text{c}_2X_2(t_k)\mathrm{e}^{\epsilon^{(2)}_{t_k}}\end{pmatrix}\eqsp, \label{eq:LV:obs:model}
\end{equation}
where $\text{c} = (\text{c}_1, \text{c}_2)^T$ is known (the observed fraction of the population) and $\lbrace \epsilon_{t_k} =(\epsilon^{(1)}_{t_k}, \epsilon^{(2)}_{t_k})\rbrace_{1\leqslant k\leqslant n}$ are i.i.d. random variables distributed as a $\mathcal{N}_2(-\text{diag}\ \Sigma/2, \Sigma)$ where $\Sigma$ is an unknown 2 $\times$ 2 covariance matrix. It is straightforward to show that for a generic $\parvec$, in the SDE defined by \eqref{eq:LV:SDE}, the drift function cannot be written (even after the Lamperti transform) as the gradient of a potential. 
Therefore, the General Poisson estimator cannot be used as an unbiased estimator of the transition density. 
Following Section \ref{sec:wald:trick}, an almost surely positive unbiased estimate of the transition density is obtained by combining the Wald's trick to the parametrix estimators of \cite{fearnhead2017continuous}.
The proposal distribution for the particle filter is again a trade off between model dynamics and the observation model (full details are given in the appendix).
 The simulated set of particles is used to obtain estimates of the true abundances given the observations, both on synthetic and real data.

\subsubsection*{Synthetic data}

\begin{figure}
\begin{center}
\includegraphics[scale = .4]{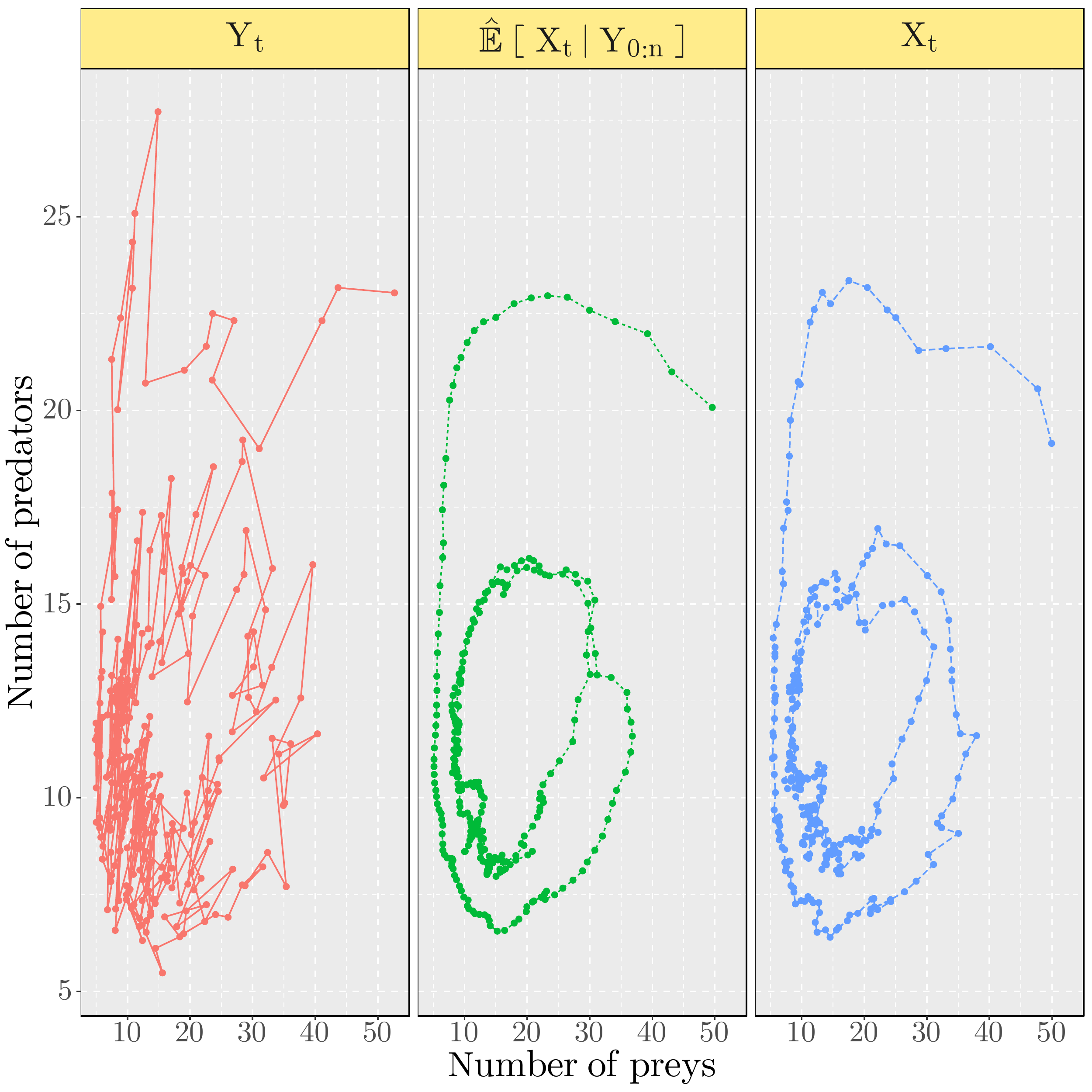}
\end{center}
\caption{\label{fig:LV:tracking} Estimated predator-prey abundances (center) in a stochastic Lotka Volterra model using our backward sampling estimate on simulated abundance indexes (left). Right panel shows the ground truth.}
\end{figure}

In a first approach,  simulated data are obtained from the model given by \eqref{eq:LV:SDE} and \eqref{eq:LV:obs:model} for a known set of parameters.
 Chosen values of $\theta$, $\Sigma$, $\text{c}_1$ and $\text{c}_2$ for the experiment are given in the appendix. 
The model is used to simulate abundances indexes $Y_0,\dots Y_{300}$ at times $t_0 = 0,\dots, t_{300} = 3$. 
The associated time series (after a division by the known constant $\text{c}$) is shown in Figure \ref{fig:LV:tracking} (left panel). 
In this experiment, the goal is to obtain an estimate of the actual predator-prey abundances given all the observed abundances indexes $Y_{0:n}$. 
Our estimate is given by the set of conditional expectations $\lbrace\pE[ X_k \vert Y_{0:n}]\rbrace_{k = 0,\dots, n}$, approximated using our backward importance sampling PaRIS smoother, which is run using the true parameters. 
Figure \ref{fig:LV:tracking} shows the estimated abundance trajectory over time. 
The proposed backward importance sampling smoother manages to estimate efficiently the actual abundance from noisy data and a model with an intractable transition density.

\subsubsection*{Hares and lynx data}

In this section, the model defined by equations \eqref{eq:LV:SDE} and \eqref{eq:LV:obs:model} is applied to the Hudson Bay company data, giving the number of hares and lynx trapped in Canada during the first 20 years of the 20th century (available in \cite{odum1971fundamentals}). 
As parameters are unknown in this case, maximum likelihood inference is performed using an EM \cite{dempster1977maximum} algorithm to obtain an estimate $\hat{\theta}$.
The E step is performed using the BIS smoother.
At each iteration, the estimator $\theta_k$ is updated by finding a parameter $\theta_{k+1}$ for which $\hat Q(\theta_{k+1},\theta_k) >  \hat Q(\theta_k,\theta_k)$, with  a gradient free evolution strategy \cite{hansen2006cma}. 
The last estimate $\hat{\theta}$ obtained with this EM algorithm  is used to  estimate the actual abundances in the model (similarly to the synthetic data case). 
Figure \ref{fig:LV:hares:lynx} shows  estimates of $\pE_{\hat{\theta}}\left[ X_k \vert Y_{0:n}\right]$ obtained with 30 independent runs of our algorithm. 
The particle smoother is implemented using $N = 200$ particles and $\tilde{N}=20$. 
The replicates show that the variance of our estimator (for a given set of observations) is much smaller than the one of the Poor Man's smoother.

\bibliographystyle{apalike}
\bibliography{backwardIS}

\appendix 

\section{Application to partially observed SDE}
\label{sec:filter:SDE}
Let $(X_t)_{t\ge 0}$ be defined as a weak solution to the following Stochastic Differential Equation (SDE) in $\mathbb{R}^d$:
\begin{equation}
\label{eq:sde}
X_0 = x_0 \quad\mbox{and}\quad \rmd X_t = \alpha_{\parvec}(X_t)\rmd t + \rmd W_t\eqsp,
\end{equation}
where $(W_t)_{t\geqslant 0}$ is a standard Brownian motion, $\alpha_{\parvec}: \Xset\to\Xset$ is the drift function
. The inference procedure presented in this paper is applied in the case where the solution to \eqref{eq:sde} is supposed to be partially observed at times $t_0 = 0,\ldots,t_n$, for a given $n\geqslant 1$, through an observation process $(Y_k)_{0\leqslant k \leqslant n}$ taking values in $\mathbb{R}^m$. For all $0\leqslant k \leqslant n$, the distribution of $Y_k$ given $(X_t)_{t\geqslant 0}$ depends on $X_k = X_{t_k}$ only and has density $\md{k;\parvec}$ with respect to $\nu$. The distribution of $X_0$ has density $\chi$ with respect to $\mu$ and for all $0\leqslant k \leqslant n-1$, the conditional distribution of $X_{k+1} $ given $(X_{t})_{0\leqslant t\leqslant k}$ has density $\hd{k+1;\parvec}(X_{k},\cdot)$ with respect to $\mu$. 
\subsection{Unbiased estimators of the transition densities}

The algorithm described above strongly relies on assumption H\ref{assum:unbiased}. 
In the context of SDEs, when $\md{k+1;\parvec}$ is available explicitly, this boils down to finding an unbiased estimate $\hdhat{k+1;\parvec}\langle \zeta\rangle(x,y)$ of $\hd{k+1;\parvec}(x,y)$ and defining
\[
\hatqg{k;\parvec}\langle \zeta\rangle(x,y) = \hdhat{k+1;\parvec}\langle \zeta\rangle(x,y)\md{k+1;\parvec}(x_{k+1},Y_{k+1})\eqsp.
\]

\subsection{General Poisson Estimators}
In \cite{olsson2011particle} and \cite{gloaguen2018online}, General Poisson Estimators (GPEs) are used to obtain an unbiased estimate of the transition density.  However, designing such estimators requires three strong assumptions \cite{beskos2006retrospective}.

\begin{enumerate}
\item The diffusion defined by \eqref{eq:sde} can be transformed into a unit diffusion through the Lamperti transform, with drift function $\tilde{\alpha}_\parvec(x)$.
\item The drift of this unit diffusion can be expressed as the gradient of a potential function, i.e., there exists a twice differentiable function $A_{\parvec}:\mathbb{R}^d \to \rset$ such $\tilde{\alpha}_{\parvec} = \nabla_x A_{\parvec}$.
\item The function $x\mapsto (\|\tilde{\alpha}_{\parvec}(x)\|^2 + \Delta A_{\parvec}(x))/2$ (where $\Delta$ denotes the Laplacian) is lower bounded.
\end{enumerate}

Assumption (1)
 is used to define a proposal distribution absolutely continuous with respect to the target which is easy to sample from. 
Assumption (2)
 is necessary to obtain a tractable Radon-Nikodym derivative between the proposal and the target distributions using the Girsanov transformation. 
 While these assumptions can be proved under mild assumptions for scalar diffusions, much stronger conditions are required in the multidimensional case \cite{ait-sahalia2008closed}. 

Let $\omega = (\omega_s)_{0 \leq s \leq t}$ be the realization of a Brownian Bridge starting at $x$ at time 0 and ending in $y$ at time $\Delta$. The distribution of $\omega$ is denoted by  $\mathbb{W}_x^{\Delta,y}$. 
Moreover, suppose that for all $\parvec\in\parspace$, $\alpha_{\parvec}$ is of a gradient form $\alpha_{\parvec} = \nabla_x A_{\parvec}$ where $A_{\parvec}:\Xset \to \rset$ is a twice continuously differentiable function. Denoting, $ \psi_\theta:~~x \mapsto  \psi_\theta(x) = (\|\alpha_{\parvec}(x)\|^2 + \Delta A_{\parvec}(x))/2$, by Girsanov theorem, for all $x, y \in \mathbb{R}^d \times \mathbb{R}^d$
\begin{equation}
\label{eq:q:girsanov}
\hd{k+1;\parvec}(x,y) = \phi_{\Delta_k}(x-y)\mathrm{exp}\left(A_{\parvec}(y) - A_{\parvec}(x)\right)\mathbb{E}_{\mathbb{W}_x^{\Delta_k,y}}\left[\mathrm{exp}\left(-\int_0^{\Delta_k} \psi_{\parvec}(\omega_s)\rmd s\right)\right]\eqsp,
\end{equation}
where $\Delta_k = t_{k+1}-t_k$, for all $a>0$, $\phi_a$ is the probability density function of a centered Gaussian random variable with variance $a$. The transition density then cannot be computed as it involves an integration over the whole path between $x$ and $y$. 
To perform the algorithm proposed in this paper, we therefore have to design a positive an unbiased estimator of $\hd{k+1;\parvec}(x,y)$. 

\paragraph{Unbiased GPE estimator for $\hd{k+1;\parvec}(x,y;\zeta)$.}

Assume that there exist random variables $\gpeLB$ and $\gpeUB$ such that for all $0\leqslant s \leqslant \Delta_k$, $\gpeLB \leqslant \psi_{\parvec}(\omega_s) \leqslant \gpeUB$. Let $\kappa$ be a random variable taking values in $\mathbb{N}$ with distribution $\mu$, $\omega = (\omega_s)_{0 \leq s \leq \Delta_k}$ be the realization of a Brownian Bridge, and $(U_j)_{1\leqslant j \leqslant \kappa}$ be independent uniform random
variables on $(0,\Delta_k)$ and $\zeta = (\kappa,\omega, U_1, \ldots , U_{\kappa})$. As
shown in \cite{fearnhead2008particle}, equation \eqref{eq:q:girsanov} leads to a positive unbiased estimator given by
\[
\hdhat{k+1;\parvec}(x,y;\zeta) = \phi_{\Delta_k}(x-y)\mathrm{exp}\left(A_{\parvec}(y) - A_{\parvec}(x) - \gpeLB\Delta_k\right)\prod_{j=1}^{\kappa}\frac{\gpeUB-\psi_{\parvec}(\omega_{{U_j}})}{\gpeUB-\gpeLB}\eqsp.
\]

\paragraph{Unbiased GPE estimator of $\nabla_{\parvec}\log\hd{k+1;\parvec}(x,y)$.}
Let's denote $\varphi_{\parvec}:~x \mapsto \psi_{\parvec}(x) - \gpeLB$. 
By \eqref{eq:q:girsanov},
\begin{multline*}
\nabla_{\parvec}\log\hd{k+1;\parvec}(x,y) = \nabla_{\parvec}A_{\parvec}(y) - \nabla_{\parvec}A_{\parvec}(x) - \nabla_{\parvec}\gpeLB \Delta_k\\
-\frac{\mathbb{E}_{\mathbb{W}_x^{\Delta_k,y}}\left[\left(\int_0^{\Delta_k} \nabla_{\parvec}\varphi_{\parvec}(\omega_s)\rmd s\right)\mathrm{exp}\left(-\int_0^{\Delta_k} \varphi_{\parvec}(\omega_s)\rmd s\right)\right]}{\mathbb{E}_{\mathbb{W}_x^{\Delta_k,y}}\left[\mathrm{exp}\left(-\int_0^{\Delta_k} \varphi_{\parvec}(\omega_s)\rmd s\right)\right]}\eqsp.
\end{multline*}
 On the other hand, the diffusion bridge $\mathbb{S}^{\Delta_k,y}_{\parvec,x}$ associated with the SDE \eqref{eq:sde} is absolutely continuous with respect to $\mathbb{W}_x^{\Delta_k,y}$ with Radon-Nikodym derivative given by
\begin{align*}
\frac{\rmd \mathbb{S}^{\Delta_k,y}_{\parvec,x}}{\rmd \mathbb{W}_x^{\Delta_k,y}}(\omega) &= \left[\hd{k+1;\parvec}(x,y)\right]^{-1}\phi_{\Delta_k}(x-y)\mathrm{exp}\left(A_{\parvec}(y) - A_{\parvec}(x) - \gpeLB\Delta_k-\int_0^{\Delta_k} \varphi_{\parvec}(\omega_s)\rmd s\right)\eqsp,\\
&=\mathbb{E}_{\mathbb{W}_x^{\Delta_k,y}}\left[\mathrm{exp}\left(-\int_0^{\Delta_k} \varphi_{\parvec}(\omega_s)\rmd s\right)\right]^{-1}\mathrm{exp}\left(-\int_0^{\Delta_k} \varphi_{\parvec}(\omega_s)\rmd s\right)\eqsp.
\end{align*}
This yields
\[
\nabla_{\parvec}\log\hd{k+1;\parvec}(x,y) = \left(\nabla_{\parvec}A_{\parvec}(y) - \nabla_{\parvec}A_{\parvec}(x) - \nabla_{\parvec}\gpeLB\Delta_k\right) - \mathbb{E}_{\mathbb{S}_{\parvec,x}^{\Delta_k,y}} \left[\int_0^{\Delta_k} \nabla_{\parvec}\varphi_{\parvec}(\omega_s)\rmd s\right]
\]
and an unbiased estimator of $\nabla_{\parvec}\log\hd{k+1;\parvec}(x,y)$ is given by
\[
\mathsf{l}_{k+1;\parvec}(x,y,\mathsf{s}^{\parvec,x,y,\Delta_k}_U) = \left(\nabla_{\parvec}A_{\parvec}(y) - \nabla_{\parvec}A_{\parvec}(x) - \nabla_{\parvec}\gpeLB\Delta_k\right) - \Delta_k\nabla_{\parvec}\varphi_{\parvec}(\mathsf{s}^{\parvec,x,y,\Delta_k}_U)\eqsp,
\]
where $U$ is uniform on $(0,1)$ and independent of $\mathsf{s}^{\parvec,x,y,\Delta_k}\sim \mathbb{S}_{\parvec,x}^{\Delta_k,y}$. In the context of GPE, $\mathsf{s}^{\parvec,x,y,\Delta_k}$ can be simulated exactly using exact algorithms for diffusion processes proposed in \cite{beskos2006retrospective}.

\subsection{Parametrix estimators}

More recently, \cite{andersson2017unbiased} and \cite{fearnhead2017continuous} proposed an algorithm which can be used under weaker assumptions. 
This parametrix algorithm draws weighted skeletons using an importance sampling mechanism for diffusion processes. 
In this case, the sampled paths are not distributed as the target process but the weighted samples produce unbiased estimates of expectations of functionals of this process. To obtain an unbiased estimator $\hdhat{k+1}\langle \zeta\rangle(x,y)$, the parametrix algorithm draws weighted skeletons at random times $s_0 = 0 < s_1<\dots<s_j $, denoted by $\left\{(x_{s_j},\mathsf{w}_{s_j})\right\}_{j\geqslant 0}$, where $x_0 = x$ and $\mathsf{w}_0=1$. 
The update times $(s_j)_{j\geqslant 0}$ are instances of an inhomogeneous Poisson process of intensity $\lambda(t)$. 
Let $(x_{s_j},\mathsf{w}_{s_j})$ be the last weighted sample and $s_{j+1}$ be the next update time of the trajectory.  While $s_{j+1}<\Delta t_{k}$, the new state is sampled using a simple Euler scheme, namely:
\begin{align*}
x_{s_{j+1}} &\eqdef x_{s_j} + \Delta s_j\alpha_{\parvec}(x_{s_j}) + (\Delta s_j)^{1/2}\sigma_{\parvec}(x_{s_j})\varepsilon_{j+1}\eqsp,
\end{align*}
where $\Delta s_j \eqdef s_{j+1}-s_j$, $\Delta t_{k} = t_{k+1} - t_{k}$ and $\varepsilon_{j+1}\sim\mathcal{N}_d(0,I_d)$. 
The proposal density associated with this procedure is denoted by $m_{j;\parvec}\left(x_{s_j},\cdot,\Delta\tau_j\right)$. 
Let $\mathcal{K}^{\parvec}$ (resp. $\mathcal{K}^{j,\parvec}_{\mathrm{prop}}$) denote the Kolmogorov forward operator of the diffusion  (resp. the Kolmogorov forward operator of the proposal distribution $m_{j;\parvec}\left(x_{s_j},\cdot,\Delta s_j
\right)$). 
The forward operators write, for any function $h:\mathbb{R}^d\rightarrow\mathbb{R}$,
$$
\mathcal{K}^{\theta}h\left(y\right) \eqdef -\sum_{i=1}^d\frac{\partial}{\partial y_i}\left\{\alpha_{\parvec,i}(y)h\left(y\right)\right\} + \sum_{i,\ell=1}^d\frac{1}{2}\frac{\partial^2}{\partial y_i\partial y_\ell}\left\{\gamma_{\parvec,i,\ell}(y)h\left(y\right)\right\}\eqsp,
$$
where $\gamma_\parvec = \sigma_\parvec \sigma_\parvec^T$.
Then, following \cite{fearnhead2017continuous}, the weight is updated by
\[
\mathsf{w}_{s_{j+1}}\eqdef\mathsf{w}_{s_j}\rho^{\lambda}_j\left(x_{s_j},x_{s_{j+1}},\Delta s_j\right)\eqsp,
\]
where
\begin{equation}
\label{CIS:eq:rho}
\rho^{\lambda}_j\left(x,y,u\right)\eqdef 1+\frac{\left(\mathcal{K}-\mathcal{K}^{j,\theta}_{\mathrm{prop}}\right)m_{j;\parvec}\left(x,z,u\right)_{|z=y}}{\lambda(u)m_{j;\parvec}\left(x,y,u\right)}\eqsp.
\end{equation}
It is worth noting that \eqref{CIS:eq:rho} can be computed using only first derivatives of $\alpha_{\parvec}$ and second derivatives of $\sigma_\parvec$. 
If $N_k$ is the number of Poisson events between $0$ and $\Delta t_{k}$, the parametrix unbiased estimate is then given by 
\[
\hdhat{k+1}\langle \zeta_k\rangle(x,y) = \mathsf{w}_{s_{N_k}}m_{k;\parvec}\left(x_{s_{N_k}},y,t_{k+1} - s_{N_k}\right) \eqsp,
\]
where $\zeta_k$ stands for all the randomness required to produce the parametrix estimator (Poisson process and Gaussian random variables).

The stability of this estimator is studied in \cite{fearnhead2017continuous} which provides $\mathrm{L}_p$ controls for the weight $\mathsf{w}_{s_{N_k}}$. 
The parametrix algorithm mentioned above is a highly flexible procedure to obtain such an unbiased estimate for a much broader class of diffusions than Poisson based estimations which require strong assumptions. 
However, as the update \eqref{CIS:eq:rho} involves the difference of two Kolmogorov operators, the parametrix estimator of the transition density may be negative, and has no reason to satisfy \eqref{eq:AR:bound}.


\end{document}